\documentclass[useAMS,usenatbib]{mn2e}
\usepackage{natbibmnfix, graphicx, times}
 \usepackage{graphicx}
 \usepackage{subfigure}
\usepackage{color,hyperref}

\usepackage{amssymb,amsmath}
\usepackage{bm}
\def\red#1 {\textcolor{red}{#1}\ }   
\def\blue#1 {\textcolor{blue}{#1}\ }   

\newcommand{\In}{\mathrm{in}}
\newcommand{\Out}{\mathrm{out}}
\newcommand{\pin}{p\text{-}\mathrm{in}}
\newcommand{\pout}{p\text{-}\mathrm{out}}
\newcommand{\inout}{\mathrm{in}\text{-}\mathrm{out}}
\newcommand{\nvec}{\mathbf{\hat{~L}}}

\newcommand{\evec}{\mathbf{e}}
\newcommand{\jvec}{\mathbf{j}}

\newcommand{\apj}{ApJ}

\newcommand{\aap}{A \& A}

\newcommand{\aj}{AJ}
\newcommand{\mnras}{MNRAS}

\newcommand{\nat}{Nature}

\title[ Survival of Planets Around Shrinking Stellar Binaries]{\Huge\bf  \fontfamily{phv}\selectfont Survival of Planets Around Shrinking Stellar Binaries}

\author[Mu\~noz \& Lai]{
\bf \Large \fontfamily{phv}\selectfont Diego J. Mu\~noz\thanks{E-mail:dmunoz@astro.cornell.edu}
and Dong Lai\\
\fontfamily{phv}\selectfont Center for Space Research, Department of Astronomy, Cornell University, Ithaca, NY 14853, USA}

\begin{document}

\maketitle

\begin{abstract}
{ \bf \fontfamily{phv}\selectfont 
The discovery of transiting circumbinary planets by the {\it Kepler} mission
suggests that planets can form efficiently around binary stars.
None of the stellar binaries currently known to host planets 
has a period shorter than 7 days, despite the large number of eclipsing binaries
found in the Kepler target list with periods shorter than a few days.
These compact binaries are believed to 
have evolved from wider orbits into their current configurations via
the so-called Lidov-Kozai migration mechanism, in which 
gravitational perturbations from a distant tertiary companion
induce large-amplitude eccentricity oscillations in the binary, followed
by orbital decay and circularization due to tidal dissipation in the stars.
Here we explore the orbital evolution of planets around 
binaries undergoing orbital decay by this mechanism. We show that  planets may survive
and become misaligned from their host binary, or may develop erratic behavior in eccentricity, resulting in
their consumption by the stars or ejection from the system as
the binary decays. Our results suggest that circumbinary planets around compact binaries
could still exist, and we offer predictions as to what their orbital configurations should be like.}
\end{abstract}

\begin{keywords}
{\fontfamily{phv}\selectfont planet dynamics and stability | close stellar binaries | multiple stellar systems}
\end{keywords}


\section*{}
\vspace{-0.25in}
To date, the {\it Kepler} spacecraft has discovered eight binary star systems
 harboring ten transiting circumbinary planets  
 \cite{doy11,wel12,oro12b,oro12a,sch13,kos13,kos14,wel14b}.
 These systems have binary periods ranging from $7.5$ to $\sim41$ days,
while the planet periods range from $\sim50$ to $\sim250$ days.
Remarkably, no transiting planets have been found around more compact
stellar binaries, those with orbital periods of $\lesssim5$~days.
Planets around such compact binaries, if orbiting in near
coplanarity, should have transited several times over the lifetime of the {\it Kepler} mission.
However,  the shortest period binary hosting a planet is Kepler-47(AB) with 7.44 days,
despite the fact that nearly 50$\%$ of the eclipsing binaries in the early quarters of Kepler 
data have periods shorter than 3 days \cite{sla11}.
Thus, the apparent absence of planets around short-period binaries
is statistically significant [e.g.,\cite{arm14}].

It is widely believed that short-period binaries ($\lesssim5$~days) 
are not primordial, but have evolved from a wider configurations
via Lidov\---Kozai (LK) cycles \cite{lid62,koz62} with tidal friction \cite{maz79,egg01,fab07}.
This ``LK+tide" mechanism requires an external tertiary companion
at high inclination to excite the inner binary eccentricity such that tidal dissipation
becomes important at pericenter, eventually leading to orbital decay and circularization. A rough transition
at an orbital period of $6$ days has been identified as the separation between ``primordial" and ``tidally evolved"
binaries \cite{fab07}. Indeed, binaries with periods shorter than this threshold are known to have very high tertiary companion
fractions [of up to $96\%$ for periods $<3$ days; see \cite{tok06}], supporting the idea that three-body interactions
have played a major role in their formation. 

In synthetic population studies \cite{fab07}, stellar binaries with periods shorter
than $\sim5$ days evolved from binaries with original periods of $\sim100$ days.
Interestingly, it is around binaries with periods $\lesssim100$ days that transiting planets have been
detected.  It is thus plausible that current compact binaries with a tertiary companion may have once 
been primordial hosts to planets like those detected by {\it Kepler}.

In this work, we study the evolution and survival of planets around stellar binaries undergoing orbital
shrinkage via the ``LK+tide" mechanism. We follow the secular evolution of the planet until binary circularization is reached
and binary separation is shrunk by an order of magnitude.  We show that
the tertiary companion can play a major role in misaligning and/or destabilizing the planet
as the binary shrinks. 
\vspace{-0.1in}
\section*{A planet inside a stellar triple}
%

\begin{figure*}
\includegraphics[width=0.93\textwidth]{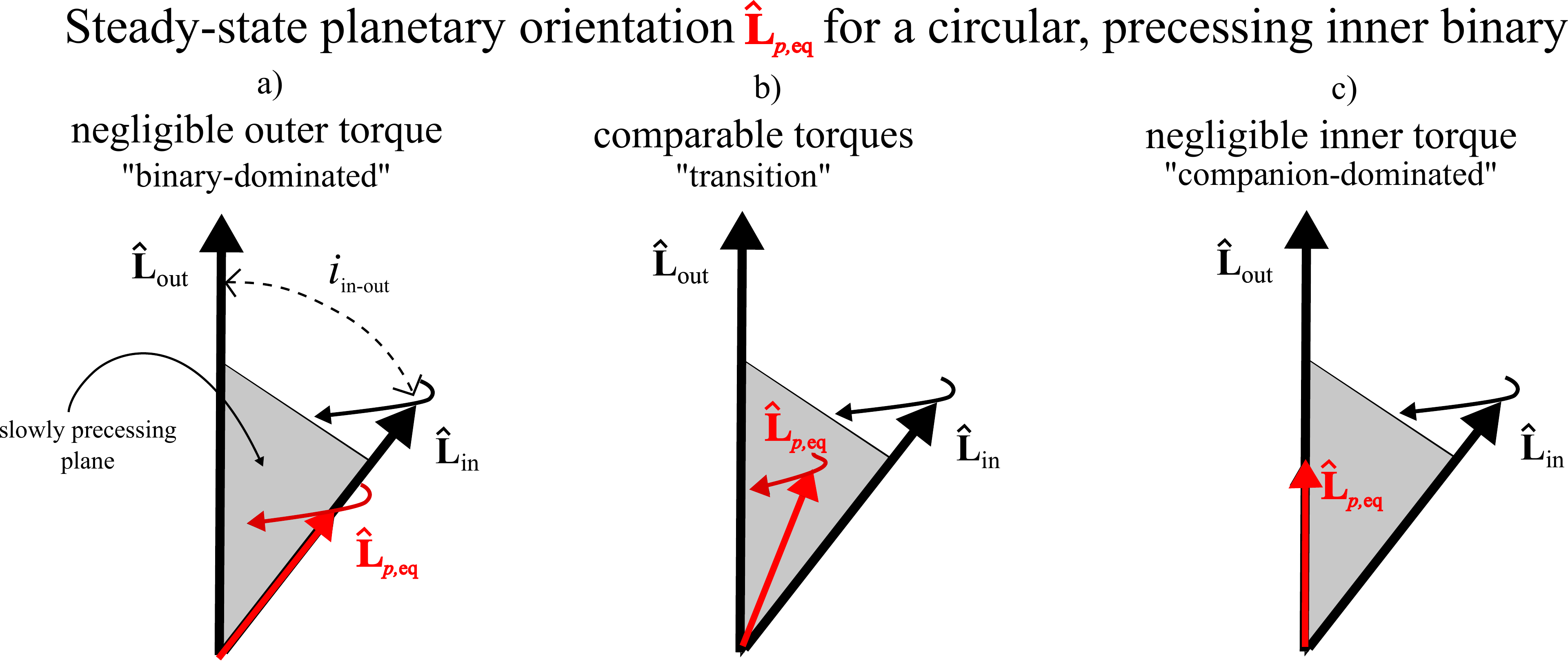}
\caption{The planet's angular momentum orientation $\nvec_p$ precesses around the equilibrium
solution $\nvec_{p,\mathrm{eq}}$ (red vector), which is obtained from balancing
the torques acting on the planet due to the inner binary and the outer companion.
The limiting cases are 
(a) $\nvec_{p,\mathrm{eq}}\parallel\nvec_\In$ (planet at small distance $a_p$ from the inner binary, 
where outer torque is negligible) and (c)  $\nvec_{p,\mathrm{eq}}\parallel\nvec_\Out$ 
(at large distance, where the inner torque is negligible), while  (b) represents the intermediate cases where
the two torques have comparable magnitudes. The continuous variation of $\nvec_{p,\mathrm{eq}}$ with $a_p$
(always coplanar with $\nvec_\In$ and $\nvec_\Out$)
going from (a) to (b) 
 to (c) defines the so-called ``Laplace surface".
 When the inner binary axis $\nvec_\In$ precesses slowly around $\nvec_\Out$ by the action of
 the outer companion, $\nvec_{p,\mathrm{eq}}$ follows adiabatically and stays coplanar
 with $\nvec_\In$ and $\nvec_\Out$. If the torque from the inner binary is slowly decreased in time,
 (e.g., due to orbital decay),  $\nvec_{p,\mathrm{eq}}$ not only will precess around $\nvec_\Out$, 
 but will also change its inclination, going smoothly from regime (a) to regime (c).
\label{fig:vector_rotation}}
\end{figure*}  

Consider a planet orbiting a 
circular stellar binary of total mass $M_\In=m_0+m_1$ and semimajor axis $a_\In$;
the binary is a member of a hierarchical triple, in
which the binary and an outer companion of mass $M_\Out$ orbit each other
with a semimajor axis $a_\Out\gg a_\In$. The secular (long term) gravitational perturbations
exerted on the planetary orbit from the quadrupole potential associated\footnote{
If the inner binary has an equal-mass ratio and the outer companion has zero
eccentricity, the octupole-order terms in the potential vanish exactly.} 
with the inner binary and that from
the outer companion cause the two vectors that determine the orbital properties of the planet, 
the angular momentum direction $\nvec_p$ and the eccentricity vector $\evec_p$, to
evolve in time. The inner binary tends to make $\nvec_p$ precess around $\nvec_\In$
, the unit vector along the inner binary's angular momentum, at a rate approximately given
by
\begin{equation}\label{eq:precession_in}
{\Omega}_{\pin}\equiv
\frac{1}{2} n_p \left(\frac{\mu_\In}{M_\In}\right)\left(\frac{a_\In}{a_p}\right)^2~~,
\end{equation}
where $a_p$ is the semi-major axis of the planet,
$n_p=\sqrt{\mathcal{G}M_\In/a_p^3}$ is the planet's mean motion frequency
(assumed to be on a circular orbit),
and $\mu_\In=m_0m_1/M_\In$ is the reduced 
mass of the inner stellar pair. Similarly, the outer companion of mass $M_\Out$
tends to make $\nvec_p$ precess around $\nvec_\Out$ at a rate approximately
given by\footnote{
Although we assume a circular outer companion here, the eccentricity of the outer
orbit $e_\Out$ can be taken into account by replace $a_\Out$ with $a_\Out\sqrt{1-e_\Out^2}~.$
},%
\begin{equation}\label{eq:precession_out}
{\Omega}_{\pout}\equiv
n_p \left(\frac{M_\Out}{M_\In}\right)\,\left(\frac{a_p}{a_\Out}\right)^3~~.
\end{equation}
In general, when the torques from the inner binary and the outer companion
are of comparable magnitude, $\nvec_p$ will precess around an intermediate
vector $\nvec_{p,\mathrm{eq}}$, which corresponds to the equilibrium solution 
(i.e., $d\nvec_p/dt=0$) of the planet's orbit under the two torques. For a general mutual
 inclination angle $i_{\inout}$ between the inner and outer orbits
 (where $\cos i_{\inout}=\nvec_\In\cdot\nvec_\Out$), the equilibrium inclination of the planet
 (the so-called ``Laplace surface"; see \cite{tre09,tam13}), can be found
 as a function of its semimajor axis, for which $\nvec_{p,\mathrm{eq}}$ is always coplanar with
$\nvec_\In$ and $\nvec_\Out$, with limiting states corresponding to
 alignment with the inner binary (i.e., $\nvec_{p,\mathrm{eq}}\parallel\nvec_\In$) at
 small $a_p$, and alignment with the outer companion (i.e., $\nvec_{p,\mathrm{eq}}\parallel\nvec_\Out$)
 at large $a_p$. The transition between these two orientations happens rapidly at
 the so-called ``Laplace radius" $r_L$, obtained by setting ${\Omega}_{p,\Out}={\Omega}_{p,\In}$,
 and is given by
\begin{equation}\label{eq:laplace_radius}
{r_L}=\left(\frac{\mu_\In}{2M_\Out}
a_\In^2 \,a_\Out^3\right)^{1/5}~~.
\end{equation}
Fig.~\ref{fig:vector_rotation} illustrates the three regimes of the planet's 
equilibrium orientation: (a) $\Omega_{\pin}\gg\Omega_{\pout}$ (binary-dominated regime, or $a_p\ll r_L$);
(b) $\Omega_{\pout}\sim\Omega_{\pin}$ (transition regime, or $a_p\sim r_L$); and (c) $\Omega_{\pout}\gg\Omega_{\pin}$
(companion-dominated regime, or $a_p\gg r_L$).

\begin{figure*}
\includegraphics[width=0.48\textwidth]{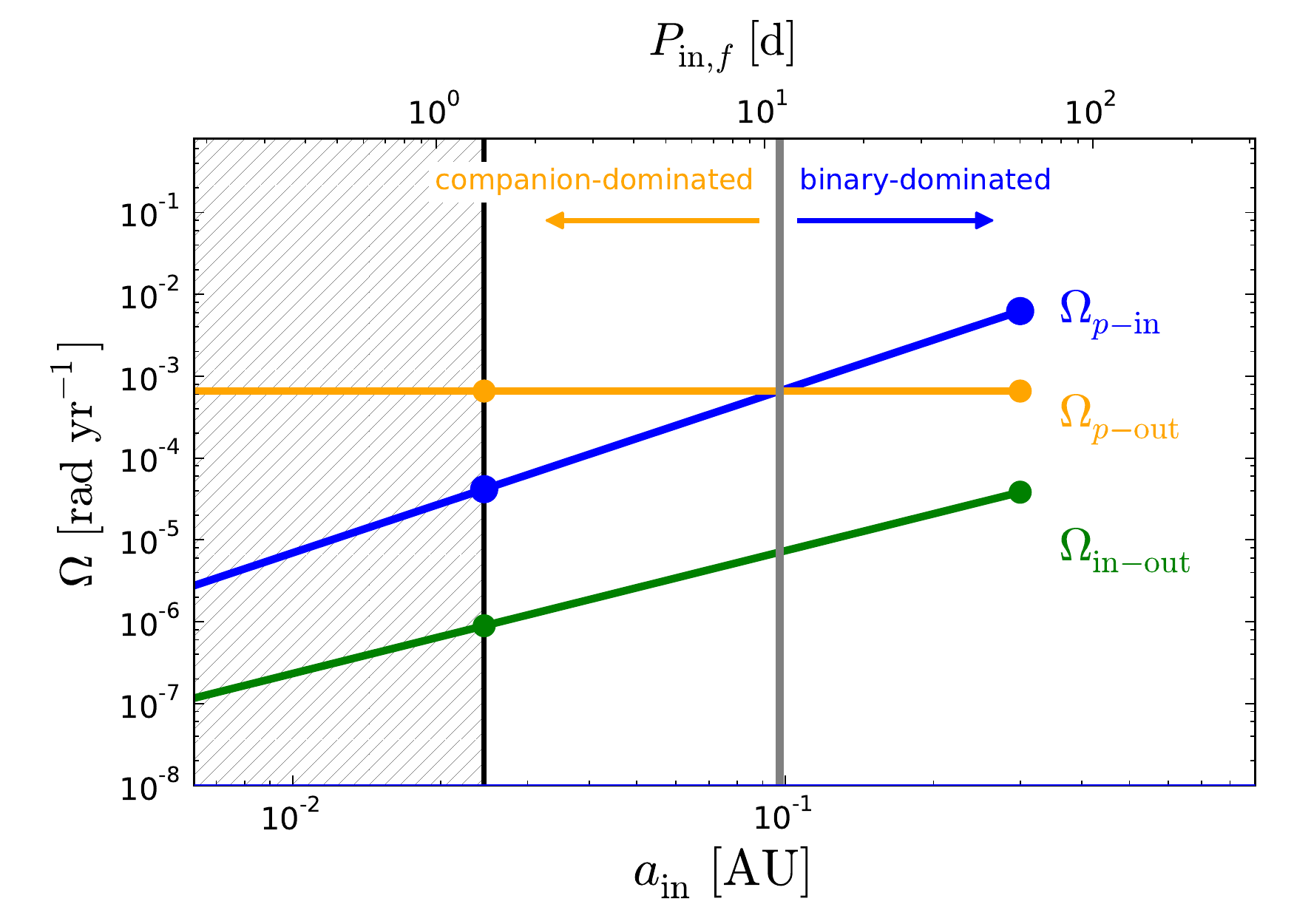}
\includegraphics[width=0.48\textwidth]{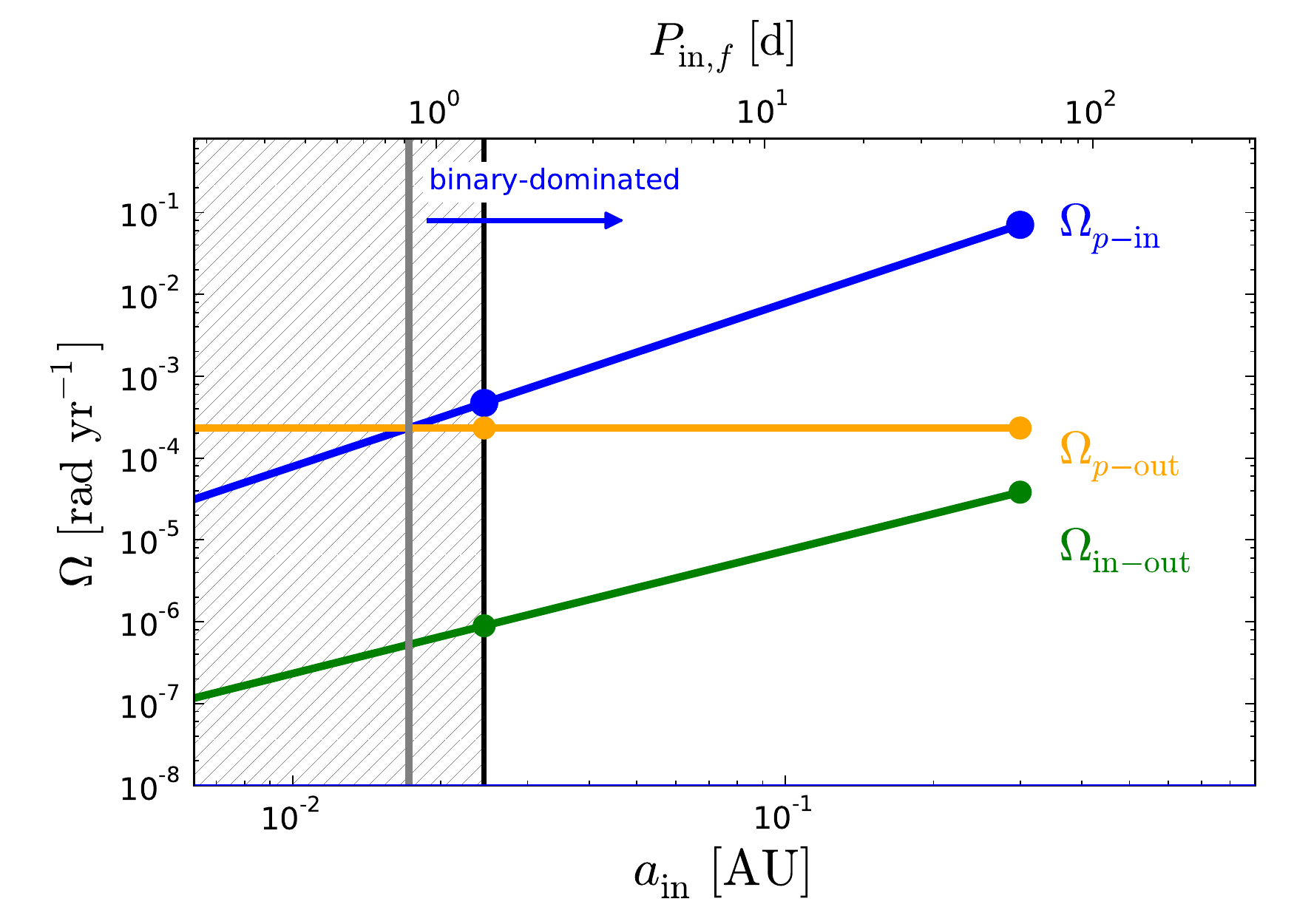}
\vspace{-0.0in}
\caption{The three relevant precession frequencies ($\Omega_{\pin}$, $\Omega_{\pout}$
and $\Omega_{\inout}$) as a function of the shrinking binary semimajor axis $a_\In$. The binary
starts at semimajor axis $a_{\In,0}=0.3$~AU
and circularizes  at $a_{\In,f}=0.024$~AU (vertical black line). The other parameters
are  $M_\In=M_\Out=1M_\odot$, $\mu_\In=0.25$, $a_\Out=30$~AU and $e_\Out=0$.
The left panel shows the case with $a_p=2$~AU, and the right panel one with $a_p=1$~AU.
On the left panel, $a_\In$ crosses
$a_{\In,L}=0.097$~AU (thick vertical gray line), during orbital decay,  then 
the planet transitions from the binary-dominated regime into the companion-dominated regime. 
On the right panel, $a_{\In,f}>a_{\In,L}=0.017$~AU) and the
planet will stay in the binary-dominated regime throughout the binary orbital decay.
Note, that in this example, $a_p=1$~AU is very close to
the initial binary, and dynamical instabilities (not captured by secular calculations) might make the survival of these
planets difficult during the early Lidov-Kozai cycles of the binary.
\label{fig:freq_ratios}}
\end{figure*}

In general, however, the vector $\nvec_\In$ is not fixed in space, but 
slowly precesses around $\nvec_\Out$, owing to the torque from the outer companion\footnote{
 Strictly speaking, both $\nvec_\In$ and $\nvec_\Out$ precess around the {\it total} angular
 momentum vector of the system; however, for the hierarchical configurations presented here,
 the outer orbit contains most of the angular momentum of the system, implying that
 $\nvec_\Out$ is approximately fixed in space.
 }. This means that the plane normal to  $\nvec_\In\times\nvec_\Out$, where the equilibrium
 orientation vector $\nvec_{p,\mathrm{eq}}$ lives, is slowly
 rotating (Fig.~\ref{fig:vector_rotation}). This rotation rate is of order
\begin{equation}\label{eq:precession_inout}
{\Omega}_{\inout}\equiv
n_\In\left(\frac{M_\Out}{M_\In}\right)\,\left(\frac{a_\In}{a_\Out}\right)^3~~,
\end{equation}
where $n_\In=\sqrt{\mathcal{G}M_\In/a_\In^3}$ is the mean motion of the inner binary.
Note that ${\Omega}_{\pout}/{\Omega}_{\inout}=(a_p/a_\In)^{3/2}\gg1$
in the companion-dominated regime, and 
 ${\Omega}_{\pin}/{\Omega}_{\inout}=({\Omega}_{\pin}/{\Omega}_{\pout})(a_p/a_\In)^{3/2}=({r_L}/{a_p})^5(a_p/a_\In)^{3/2}\gg1$
in the binary-dominated regime. This means that the precession of $\nvec_\In$
 is always slow enough for $\nvec_p$ to adiabatically follow.
In other words, the classical Laplace equilibrium
formalism remains valid in the frame  corotating with $\nvec_\In$, and the three
vectors $\nvec_\In$, $\nvec_\Out$ and $\nvec_{p,\mathrm{eq}}$ remain coplanar at all times. Since the evolution is
adiabatic, if $\nvec_p$ starts parallel to $\nvec_{p,\mathrm{eq}}$, it will remain parallel to the evolving $\nvec_{p,\mathrm{eq}}$
at later times, provided that this equilibrium orientation is a stable solution \cite{tre09}.

As studied by \cite{tre09}, when $i_{\inout}>69^\circ$,
circular orbits on the Laplace surface are unstable to linear perturbations
in the planet's eccentricity vector $\evec_p$ vector for a range of $a_p$ around $r_L$.
This instability manifests itself as an exponential growth of $e_p$,
until non-linear effects come into play, resulting
in erratic behavior in both inclination and eccentricity. This means that above
this critical value of $i_{\inout}$, planets cannot be
placed at $a_p\sim r_L$, since the resulting high eccentricities could bring them
too close to the binary, at which point they may
collide with the central stars or be ejected from the system [e.g., \cite{hol99}].


Now consider what will happen to the planet's orbit as the inner binary undergoes 
orbital decay.
For simplicity, let us
assume that the binary remains circular during this process, and that the angle
$i_{\inout}$ remains unchanged. Since orbital decay takes place over
a time scale $t_\mathrm{decay}$ much longer than the other relevant time scales
($1/{\Omega}_{\pin}$, $1/{\Omega}_{\pout}$ and $1/{\Omega}_{\inout}$),
the system will evolve adiabatically. Thus, if the planet initially resides
in the binary-dominated regime (${\Omega}_{\pin}\gg{\Omega}_{\pout}$, $a_p\ll r_L$),
and lives on the Laplace surface ($\nvec_p\parallel\nvec_{p,\mathrm{eq}}$), it will
transition to the companion-dominated regime (${\Omega}_{\pin}\ll{\Omega}_{\pout}$, $a_p\gg r_L$)
through the intermediate stage (${\Omega}_{\pin}\sim{\Omega}_{\pout}$),
as the inner binary's semi-major axis $a_\In$ decreases. For a given value of $a_p$,
the transition occurs when $a_\In$ passes
through a critical (``Laplace") value,
 \begin{equation}
 a_{\In,L}\equiv 0.017 ~\bigg(\frac{M_\Out}{4\mu_\In}\bigg)^{1/2}\bigg(\frac{a_p}{1~\mathrm{AU}}\bigg)^{5/2}
 \bigg(\frac{a_\Out}{30~\mathrm{AU}}\bigg)^{-3/2}~\mathrm{AU}
 \end{equation}
 obtained by replacing $r_L=a_p$ in Eq.~\ref{eq:laplace_radius} and solving for $a_\In$.
 If the transition region ($a_p\sim r_L$) is stable, we expect the planet's orbit to evolve smoothly
 following the Laplace surface [i.e., (a)$\rightarrow$(b)$\rightarrow$(c) in Fig.~\ref{fig:vector_rotation}].
 For $i_{\inout}>69^\circ$, however, the planet will encounter an instability when $a_p\sim r_L$, and
 may undergo erratic evolution, which may result in the planet being destroyed or ejected.

In the ``LK+tide" scenario for the formation of compact binaries,
the final inner binary separation $a_{\In,f}$ depends on the properties
of the outer companion ($M_\Out$, $a_\Out$ and the initial inclination $i_{\inout}$)
as well as on the short-range force effects between the inner binary members \cite{fab07}.
Thus, for a given stellar triple configuration, the inner binary may or may not reach down to
$a_{\In,L}$,  depending on the value of $a_p$ (see Fig.~\ref{fig:freq_ratios}). If
$a_{\In,f}>a_{\In,L}$, or equivalently, if
 \begin{equation}
 a_p < 1.26 ~\bigg(\frac{M_\Out}{4\mu_\In}\bigg)^{-1/5}
 \bigg(\frac{a_\Out}{30~\mathrm{AU}}\bigg)^{3/5}
 \bigg(\frac{a_{\In,f}}{0.03~\mathrm{AU}}\bigg)^{2/5}~\mathrm{AU}~~,
 \end{equation}
the planet will never cross the intermediate regime ($a_p\sim r_L$),
and it will thus remain ``safe" (stable), regardless of the inclination $i_{\inout}$,
surviving the orbital decay of its host binary.  

\vspace{-0.1in}
\section*{Evolution of planetary orbits around binaries undergoing Lidov-Kozai cycles
with tidal friction}
 The greatest caveat to the application of classical Laplace equilibrium is that the inner binary
does not remain circular during orbital decay. Indeed, in the ``LK+tide" mechanism \cite{egg01,fab07}
the inner binary exhibits large oscillations in inclination {\it and} eccentricity under the influence 
of the external stellar companion. Thus the binary axis $\nvec_\In$ not only precesses around
$\nvec_\Out$, but also undergoes nutation. The variation of the inner binary's
eccentricity vector $\evec_\In$ also affects the torque on the circumbinary planet.

To track the evolution of the planet's orbit during the LK oscillations  and
orbital decay of the inner binary, we solve numerically the secular equations of 
\footnote{
The secular equations of motion govern the evolution of the
orbital elements instead of the position and velocity of individual bodies.}
of the planet's eccentricity vector $\evec_p$ and angular momentum vector axis $\nvec_p$
(see Supplementary Information),
along with the evolution equations of the stellar triple. We use the formalism of
\cite{egg98} to follow the inner binary's orbit and parametrize the stellar tidal dissipation
rate using the weak friction model with constant tidal lag time. In the following,
we focus on a few representative examples and discuss the general behavior for the
evolution of the four-body system.

\begin{figure}
\centering
\includegraphics[width=0.47\textwidth]{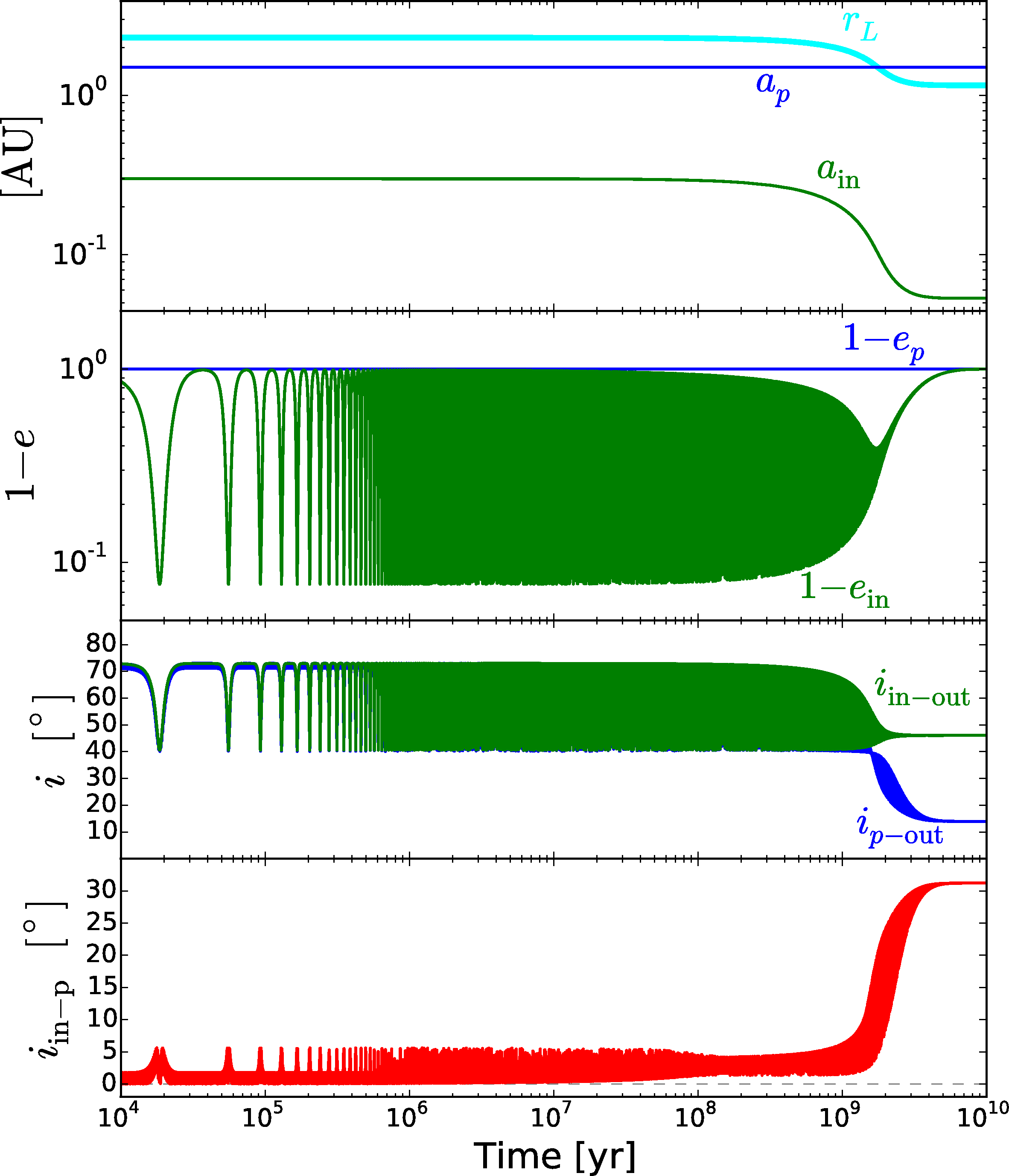}
\vspace{+0.1in}
\caption{An example of the coupled evolution of an inner binary within a stellar triple 
($m_0=m_1=0.5M_\odot$, $M_\Out=1M_\odot$,  $a_\Out=18$~AU and initial $a_{\In,0}=0.3$~AU and $i_{\inout,0}=73^\circ$)
plus a planet with semimajor axis $a_p=1.5$~AU.
The different panels show: (top) binary semimajor axis $a_\In$ (green) and planet semimajor axis $a_p$ (blue)
and the Laplace radius $r_L$ (cyan); (top middle) eccentricity
of the binary $e_\In$ (green) and eccentricity of the planer $e_p$ (blue); (top bottom) inclination of the binary $i_{\inout}$ (green)
and inclination of the planet $i_{\pout}$ (blue) with respect to the outer companion; and (bottom) mutual inclination between
the planet and the binary $i_{\pin}$. The eccentricity and inclination of the binary exhibit LK cycles
for about $10\%$ of the integration time, until short-range forces 
arrest these oscillations (freezing $e_\In$ at high values), 
at which point a slow phase of orbital decay takes place. 
The planet starts in the binary-dominated regime ($a_p/r_{L,0}= 0.65$). Its inclination $i_{\pout}$
follows closely that of the inner binary $i_{\inout}$ until $r_L$ crosses $a_p$
 (note that the definition of $r_L$ in Eq.~\ref{eq:laplace_radius} does not take into account 
the eccentricity of the inner binary $e_\In$), at which point these two inclination angles decouple from
each other. The planet ends in the companion dominated regime ($a_p/r_{L,f}= 1.3$), and
its inclination with respect to the binary $i_{\pin}$ eventually settles into a constant value 
of $\sim32^\circ$.
 \label{fig:smooth_transition}}
\end{figure}

Fig.~\ref{fig:smooth_transition} depicts a system where the stellar triple
has parameters $m_0=m_1=0.5M_\odot$, $M_\Out=1M_\odot$, $a_\Out=18$~AU and $e_\Out=0$
and initial values $a_{\In,0}=0.3$~AU and $i_{{\inout},0}=73^\circ$, and where
the planet is initialized on a circular orbit at $a_p=1.5$~AU with $\nvec_p$ aligned with $\nvec_\In$.
{The parameters for the inner binary and the planet are chosen to roughly correspond to
the} {discovered {\it Kepler} systems. The parameters for the outer orbit are chosen to ensure that
LK cycles are not suppressed by short-range forces and to guarantee the efficient orbital decay of the inner binary
\cite{fab07}. In our calculations, the octupole term in the potential has been ignored in the evolution equations of the planet and
the inner binary, a justified simplification since $m_0=m_1$ and $e_\Out=0$.}
 The inner binary experiences LK oscillations and circularizes within a Hubble time
provided that enough tidal dissipation is present in the stars. 
The final (circularization) semimajor axis is  $a_{\In,f}=0.053$~AU
(corresponding to an orbital period of 4.5~days). 
\begin{figure}
\centering
\includegraphics[width=0.45\textwidth]{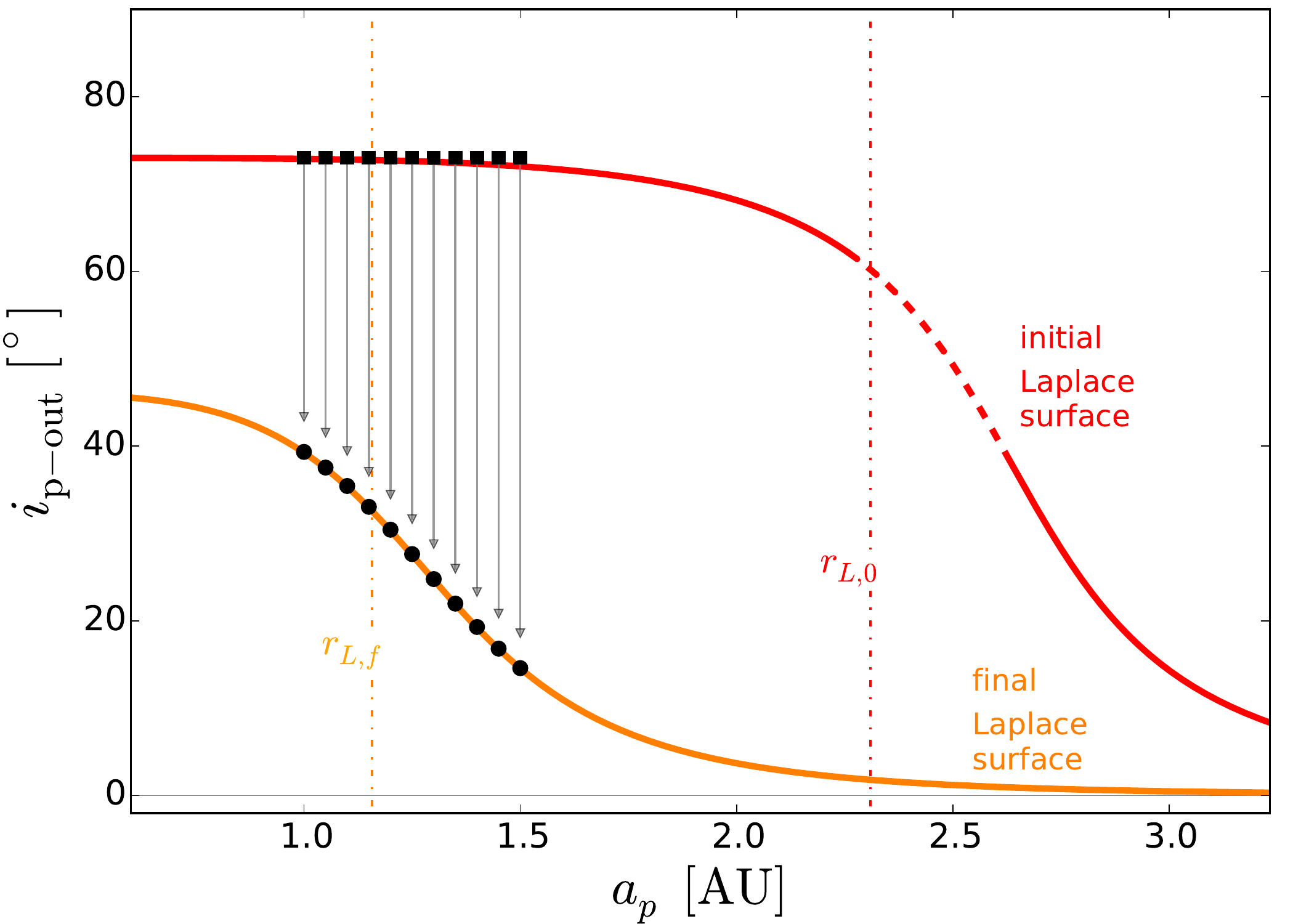}
\caption{Classical Laplace equilibrium surface (valid for $e_\In=0$)
at the beginning (red curve) and after circularization of the inner binary (orange curve) 
 for the triple configuration of Fig.~\ref{fig:smooth_transition} 
 ($m_0=m_1=0.5M_\odot$, $M_\Out=1M_\odot$, 
 $a_{\In,0}=0.3$~AU, $i_{\inout,0}=73^\circ$, $a_{\In,f}=0.053$~AU, $i_{\inout,f}=46.1^\circ$ and $a_\Out=18$~AU).
 The dotted portion of the red line indicates the range of radii at which the equilibrium surface
 is unstable \citep{tre09}. 
 The final Laplace surface is stable for all $a_p$ since $i_{{\pout},f}<60^\circ$.
 The vertical dash-dotted lines
 indicate the Laplace radii at the beginning ($r_{L,0}$) and end ($r_{L,f}$) of the binary orbital evolution.
 For the different values of $a_p$, vertical arrows connect the initial and final states, representing the 
 evolution of the planet's inclination obtained from the numerical calculations.
 In each case, the planet orientation is initially aligned with the local Laplace surface
 (or approximately aligned with the binary for $a_p\lesssim1.5$~AU); after the inner binary has
 decayed and circularized, the planet inclination settles into a value coincident with
 final Laplace surface. Note that, for illustrative purposes, we include values of $a_p$ down to 1~AU;
 however, dynamical stability dictates that only planets outside  $a_p\sim 4 a_\In\sim 1.2$~AU \citep[when $e\In\sim1$;][]{hol99}
 should survive the early LK cycles of the inner binary.\label{fig:classical_laplace}}
  \vspace{-0.1in}
\end{figure}
In this example, the planet initially resides in the binary-dominated regime, with 
${{\Omega}_{\pin}}/{{\Omega}_{\pout}}\approx 65\left({a_p}/{\mathrm{AU}}\right)^{-5}\approx8.6$,
and $a_p/r_{L,0}\approx0.65$. After the inner binary has circularized, the planet lies
in the companion-dominated regime, with ${{\Omega}_{\pin}}/{{\Omega}_{\pout}}\approx0.27$ and
$a_p/r_{L,f}\approx1.3$. We see that the planet remains on a circular orbit throughout its
entire evolution, despite the large variations in $e_\In$ during the LK cycles. The longitude
of nodes of the planet (not shown in the figure) closely follows that of the inner binary
during the early LK cycles and after circularization, implying that for a large fraction of the
time $\nvec_p$ is coplanar with $\nvec_\In$ and $\nvec_\Out$. The planet's inclination $i_{\pout}$
also follows  the binary inclination $i_{\inout}$ during the early stage of the LK cycles (third panel from top), but it 
decouples from the inner binary after $a_\In$ has started decreasing. At the end of the integration,
when the binary has circularized, the
binary and planet are misaligned by $32^\circ$ (fourth panel) and the planet inclination has settled onto
a steady-state value. This final value $i_{\pout}\simeq14^\circ$ agrees with the equilibrium value of
the end-state Laplace surface (with $a_{\In,f}=0.053$~AU and $i_{{\inout},f}=46^\circ$) evaluated at $a_p=1.5$~AU.
\begin{figure*}
\includegraphics[width=0.48\textwidth]{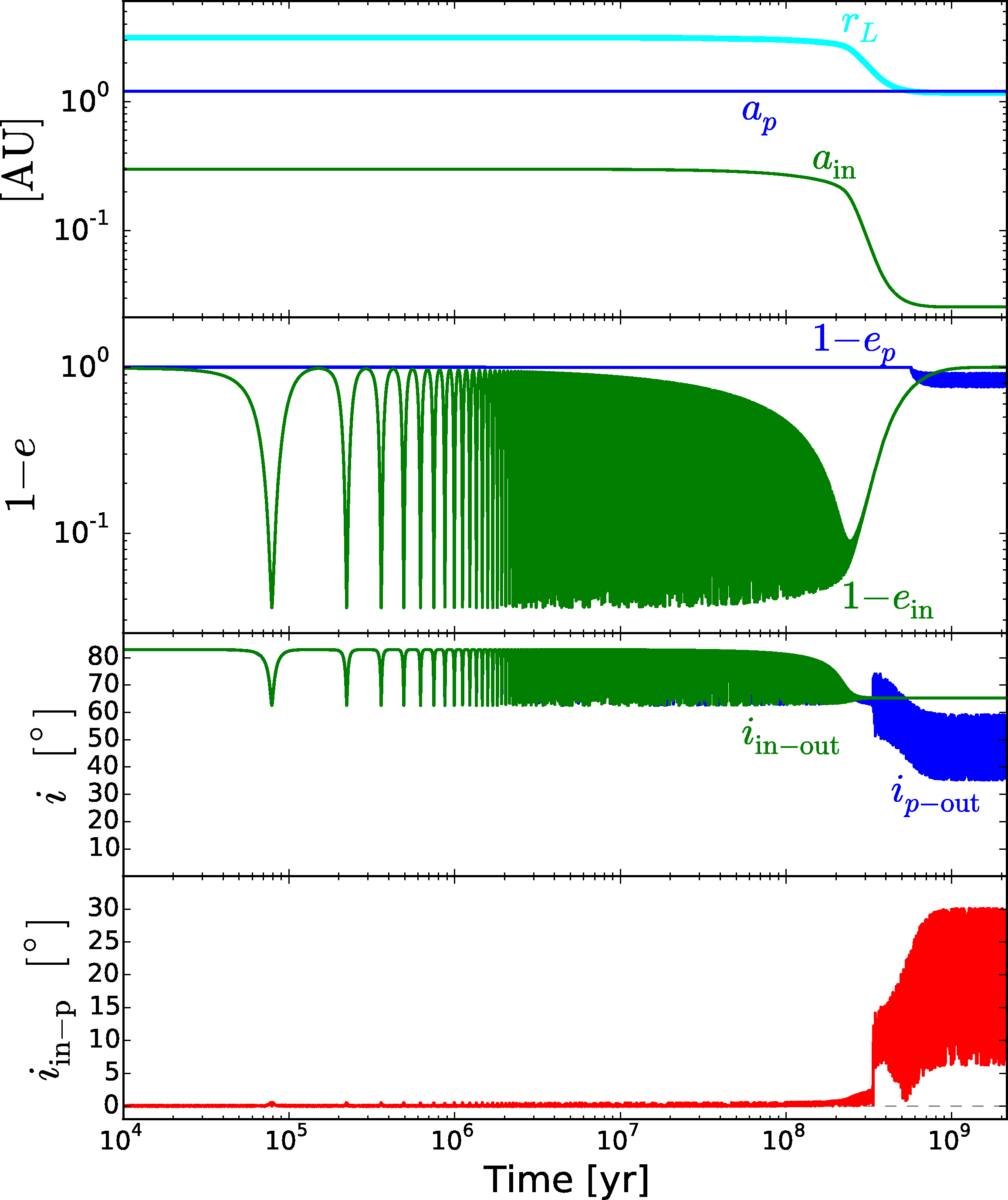}
\includegraphics[width=0.48\textwidth]{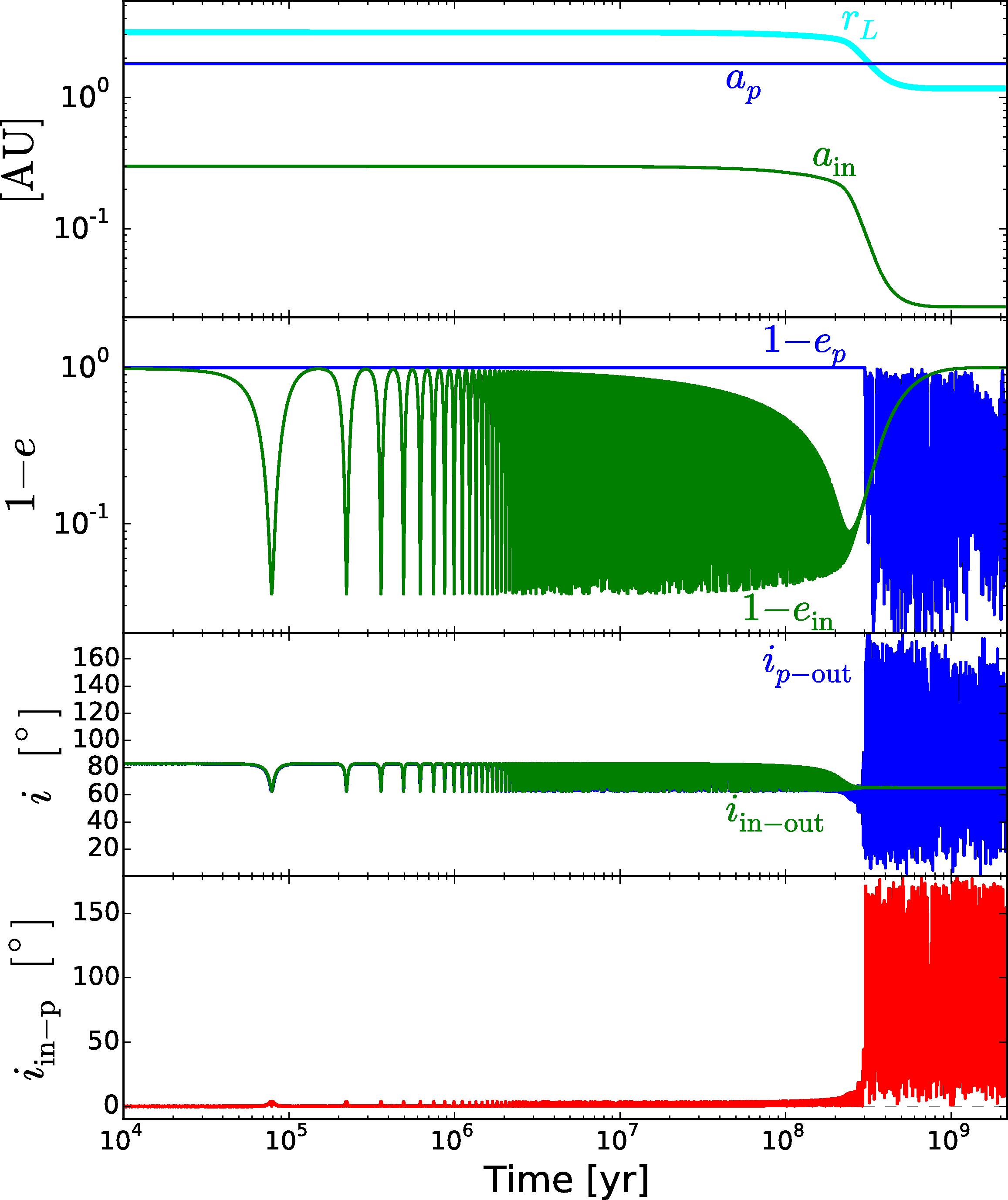}
\vspace{0.1in}
\caption{Similar to Fig.~\ref{fig:smooth_transition}, but for a triple system that is initialized
at a higher inclination $i_{\inout,0}=83^\circ$. The other parameters for the tare $a_{\In,0}=0.3$~AU and
$a_\Out=30$~AU, with the same stellar masses as in Fig.~\ref{fig:smooth_transition}.
Two examples are shown: $a_p=1.2$~AU (left panels) and $a_p=1.8$~AU (right panels), both exhibiting
quite different planetary evolution compared to Fig.~\ref{fig:smooth_transition}. In the case
of $a_p=1.2$~AU, the planet starts with $a_p/r_{L,0}=0.38$ and ends with $a_p/r_{L,f}=1.02$;
in the case of $a_p=1.8$~AU, the planet  starts with $a_p/r_{L,0}=0.57$ and ends with $a_p/r_{L,f}=1.54$.
\label{fig:erratic_transition}}
\vspace{-0.1in}
\end{figure*}

We have carried out calculations for a range of values of $a_p$ for the same
stellar triple configuration of Fig.~\ref{fig:smooth_transition}. The results of these
calculations are summarized in Fig.~\ref{fig:classical_laplace}, which shows
the Laplace surfaces at the beginning and at the end of the evolution, when
the inner binary is circular. In each case, the planet is initially aligned with the equilibrium
orientation $\nvec_{p,\mathrm{eq}}$, which is in near alignment with the inner
binary for $a_p\lesssim1.5$~AU. We find that the planet's inclination evolves smoothly
for all these cases as the binary experiences LK oscillations and orbital decay.
Despite the complexity of the ``intermediate" states, in which the binary develops
large eccentricities and the standard Laplace equilibrium is not well defined, we find that in the end,
the planet's inclination always lands on the final Laplace surface. Thus, these planets
survive the orbital decay of the inner binary, but become inclined respect to it by an angle
given by $i_{{\pin},f}=i_{{\inout},f}-i_{{\pout},f}$ (with $i_{{\inout},f}\approx46^\circ$ for
the parameters adopted in Figs.~\ref{fig:smooth_transition} and~\ref{fig:classical_laplace}),
where $i_{{\pout},f}$ matches the equilibrium inclination
of the final Laplace surface. Since the Laplace equilibrium inclination angle decreases with increasing
$a_p$, we predict that the angle $i_{\pin}$ of the planets that survive will increase monotonically with increasing $a_p$.

As noted before, when the mutual inclination $i_{\inout}$ between the inner circular binary
and the external companion is greater than $69^\circ$, a portion of the Laplace surface is
unstable \cite{tre09}.  In principle, a circumbinary planet may suffer a similar instability
as a binary with large initial $i_{\inout}$ undergoes ``LK+tide" orbital decay.
In Fig.~\ref{fig:erratic_transition}, we show
 two examples ($a_p=1.2$~AU and $a_p=1.8$~AU for the left and right
panels, respectively) of planets within a stellar triple with  $a_{\Out}=30$~AU, 
$a_{\In,0}=0.3$~AU, and $i_{{\inout},0}=83^\circ$ (the other parameters
are the same as in Fig.~\ref{fig:smooth_transition}). At this initial inclination, the inner binary
attains very high eccentricities and can circularize very efficiently (alternatively, it 
requires relatively small tidal dissipation in the stars to circularize within a Hubble time).
The final binary separation is $a_{\In,f}=2.55\times10^{-2}$~AU (period of 1.5~d) and the 
inclination angle freezes out at $i_{{\inout},f}=65.3^\circ$.
The behavior of the planets is markedly different from the one depicted in
Fig.~\ref{fig:smooth_transition}. For a planet located at $a_p=1.2$~AU 
(left panels of Fig.~\ref{fig:erratic_transition}), the inclination angle $i_{\pout}$ does
not evolve smoothly as the inner binary decays, but
 suffers a jump as $r_L$ crosses $a_p$, subsequently oscillating around 
a reference angle. Moreover, the orbital eccentricity rapidly grows until it starts
oscillating around a mean value of $\langle e_p\rangle \sim0.16$, maintaining from then on a steady-state behavior.
For a planet at $a_p=1.8$~AU (right panels), the orbital evolution is even more complex.
In this case, the exponential growth in eccentricity does not saturate at a moderate value. Instead,
$e_p$ reaches values close to 1. The erratic evolution in $e_p$ is accompanied by a similar
behavior in the planet's inclination $i_{\pout}$. Instead of oscillating around a mean (equilibrium)
value,  $i_{\pout}$ covers the entire range $(0^\circ,180^\circ)$.
The high planet eccentricities reached in this case make it very unlikely for the planet to 
survive the orbital decay of the inner binary. Such high eccentricities 
will inevitably bring the planet too close to the inner binary, a region that is known to
be unstable \cite{hol99,mud06}.
In this case, ejections from the system or physical collisions with the central stars are to be expected.
\begin{figure}
\centering
\includegraphics[width=0.48\textwidth]{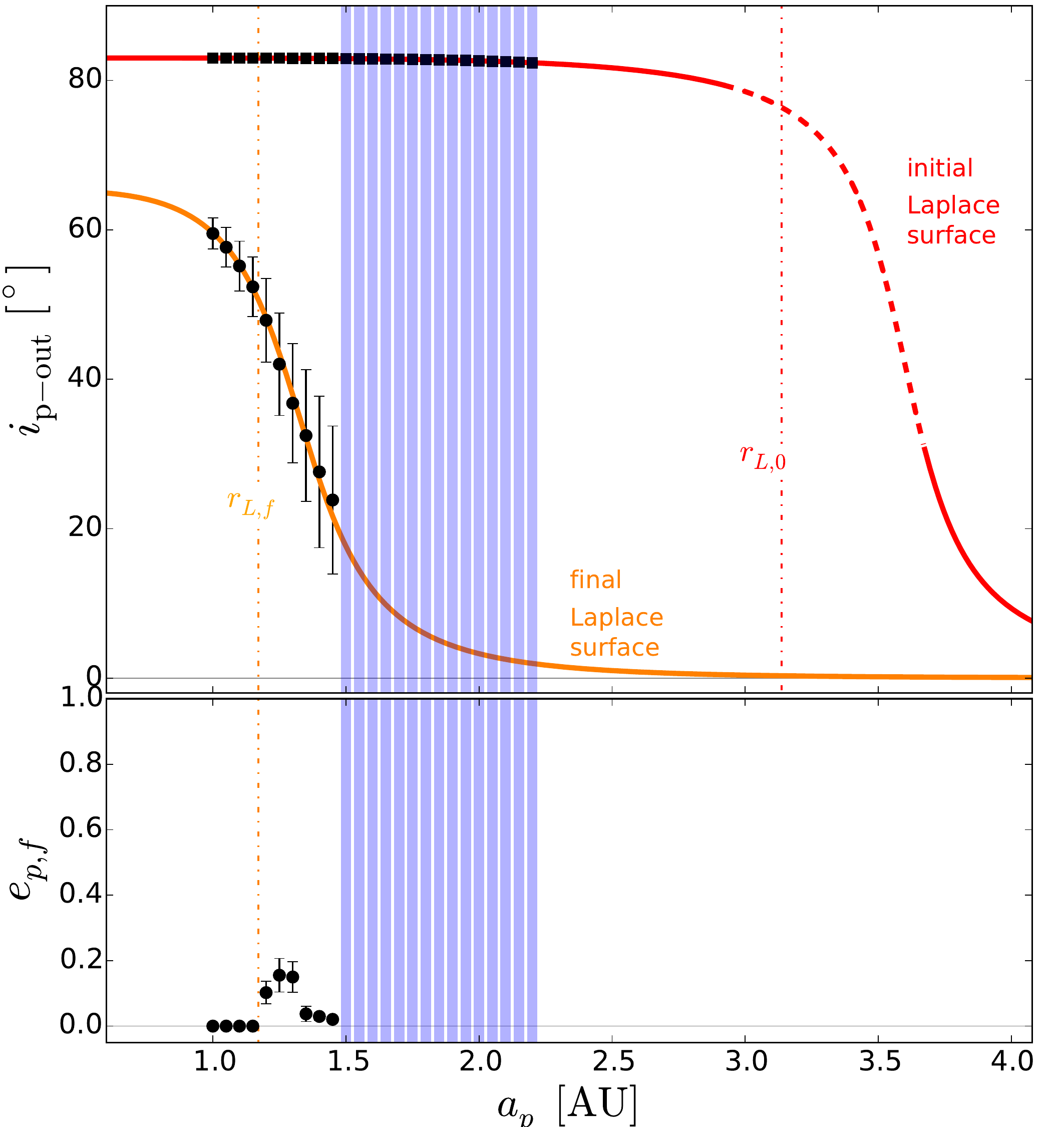}
\caption{Top panel: similar to Fig.~\ref{fig:classical_laplace}, but for the triple configuration
shown in Fig.~\ref{fig:erratic_transition} ($a_{\In,0}=0.3$~AU, $i_{\inout,0}=83^\circ$, $a_{\In,f}=0.026$~AU,
 $i_{\inout,f}=65.3^\circ$). Bottom panel: the corresponding {\it final} eccentricity of each planet.
 Error bars specify the oscillation amplitude around the mean value.
 Purple symbols denote planet orbits that were excited into eccentric states.
 Wide blue bands (for $a_p\gtrsim1.5$~AU) denote planet orbits
 with erratic behavior in inclination, covering the entire range $[0^\circ,180^\circ)$, at {\it any}
 point during their evolution. Note that, for some values of $a_p$, planets do
 reach regular values in eccentricity and inclination even {\it after} having experienced erratic evolution during
 a finite period of time before binary circularization; such cases are still depicted by blue bands, since their
 survival is deemed unlikely (see Fig~S2). 
\vspace{-0.1in}
 \label{fig:classical_laplace2}}
\end{figure}

In Fig.~\ref{fig:classical_laplace2}, we show the initial and final inclinations (top panel),
and the respective final eccentricities (bottom panel),
computed for a set of values of $a_p$ using
 the same stellar triple configuration of Fig.~\ref{fig:erratic_transition}.
As in Fig.~\ref{fig:classical_laplace}, 
these results are shown together with the Laplace equilibrium surface solutions
for the initial state ($a_{\In,0}=0.3$~AU and $i_{\inout,0}=83^\circ$)
and the final state ($a_{\In,f}=0.026$~AU and $i_{\inout,f}=65.3^\circ$).
Unlike Fig.~\ref{fig:classical_laplace}, we find that, depending on $a_p$, 
planet orbits do not always stay circular, and their inclinations $i_{\pout}$ do 
not always land exactly on the final Laplace surface. For $a_p\lesssim r_{L,f}$,
planets end up very close (on average) to the final Laplace surface (while exhibiting
some minor oscillations around it), and maintain a negligible eccentricity.  
For $a_p\gtrsim r_{L,f}$, planets suffer a small kick in eccentricity as they
cross the ``transition" regime ($a_p=r_L$), and their inclinations oscillate with
significant amplitude around a mean value that is close, but not necessarily equal to,
the one given by Laplace equilibrium (see Fig.~\ref{fig:erratic_transition}, left panel).
At even larger $a_p$ ($\gtrsim1.5$~AU), we find that the evolution of the planet is no
longer regular (see Fig.~\ref{fig:erratic_transition}, right panel): both $e_p$ and $i_{\pout}$
undergo large-amplitude, erratic variations ($e_p\simeq0$ - $1$ and $I_{\pout}\simeq0^\circ$-$180^\circ$).
Indeed, for large values of $a_p$,  the planet's evolution is most likely chaotic,
since the results depend
sensitively on the initial conditions (see Supplementary Information and Fig.~S2).
Erratic evolution (in eccentricity and inclination) may last indefinitely
 or  may end before circularization of the inner binary has completed, in which case 
 planets can exit the erratic phase at a random inclination
(including angles $>90^\circ$). In either case, these planets, having experienced erratic, large-amplitude
variations of $e_p$, are likely to be ejected from the system or to collide with the binary stars.

\begin{figure*}
\centering
\includegraphics[width=0.45\textwidth]{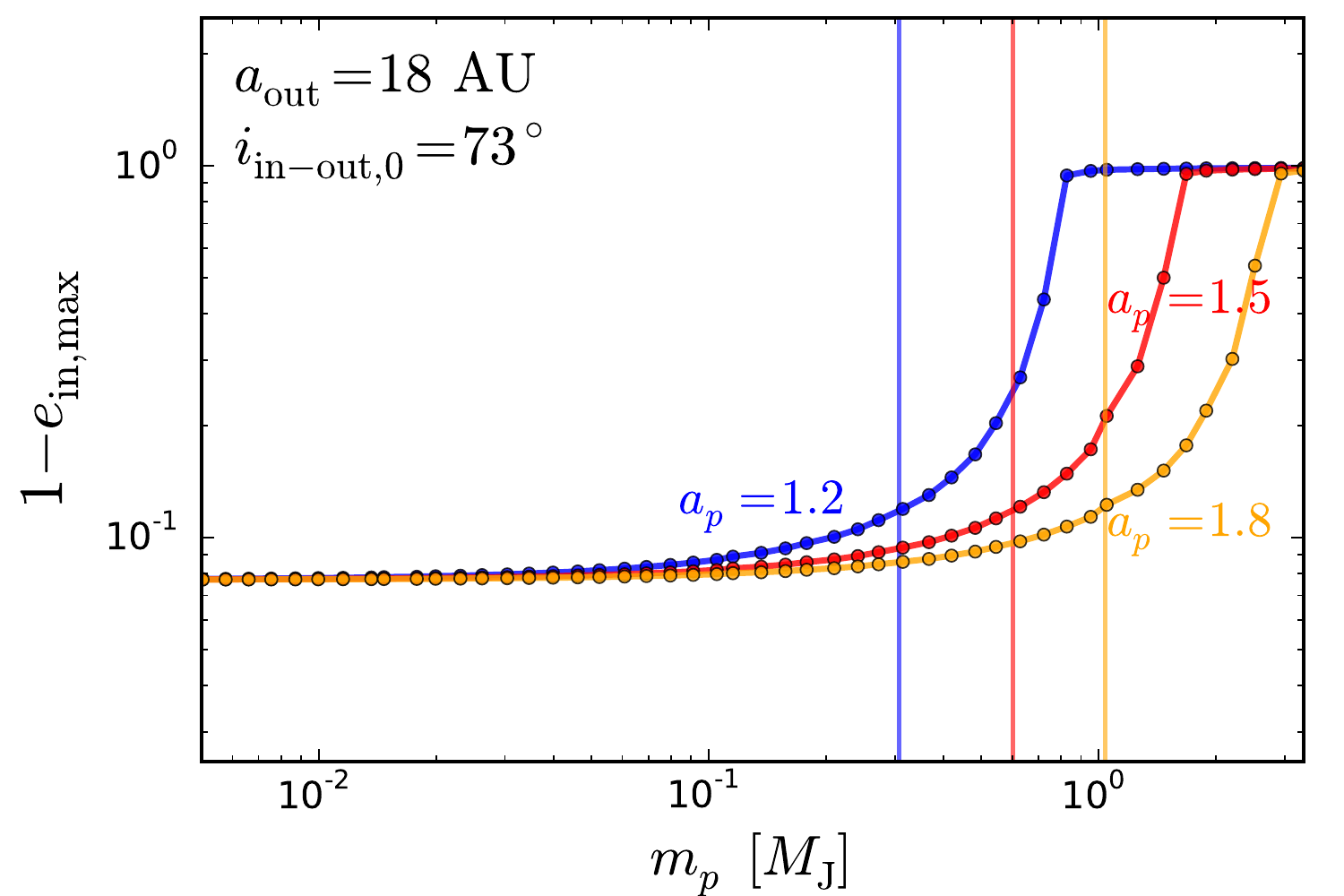}
\includegraphics[width=0.45\textwidth]{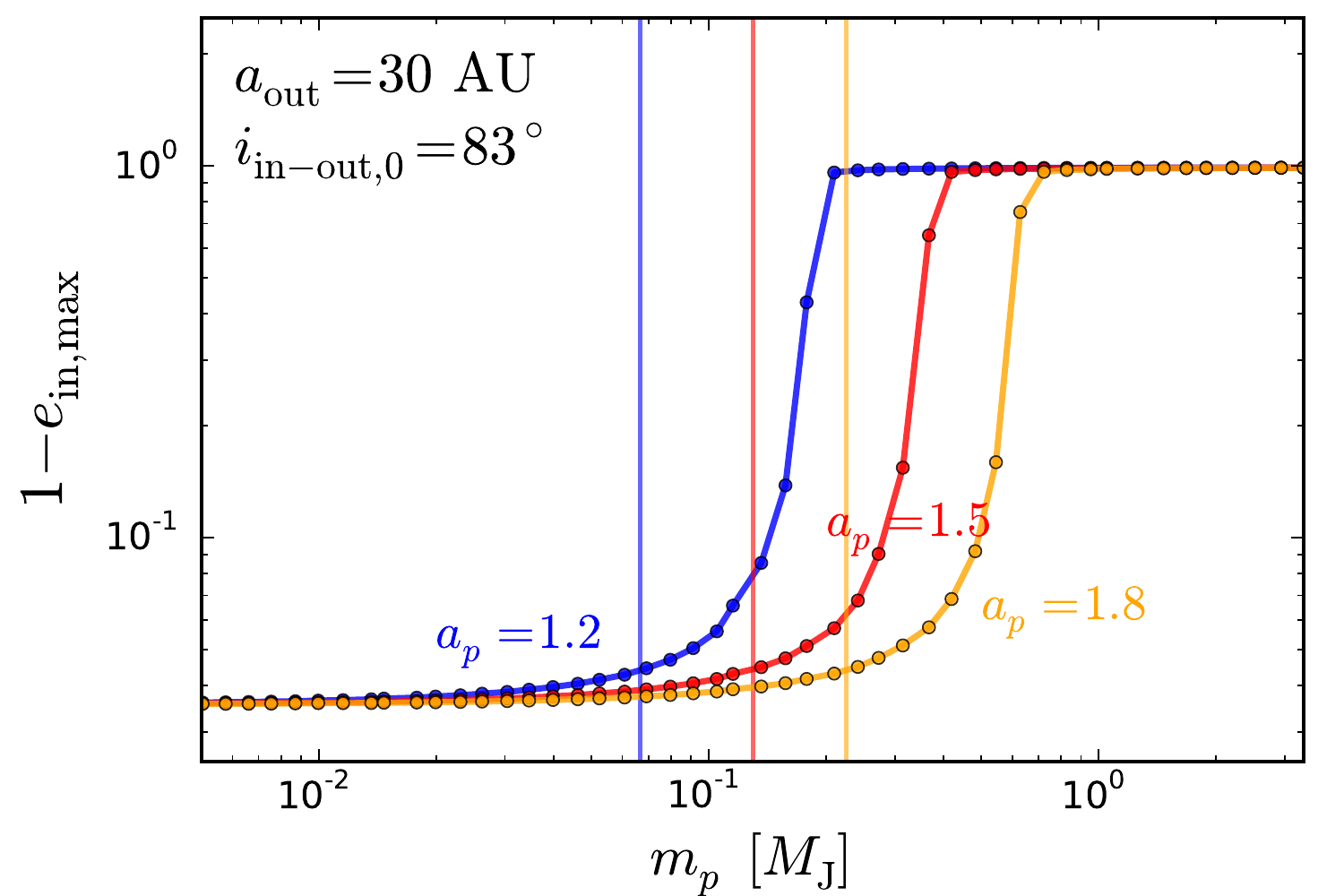}
\caption{Left: maximum eccentricity $e_{\In,\mathrm{max}}$ of the inner binary in the triple
configuration of Fig.~\ref{fig:classical_laplace} achieved during
the LK cycles as a function of planet mass $m_p$
for three different values of the planet  semimajor axis $a_p$: 1.2 AU (blue), 1.5 AU (red) and 1.8 AU (orange).
Vertical lines denote
the value of $m_{p,\mathrm{crit}}$ (Eq.~\ref{eq:mcrit}) for each of the different values of $a_p$.
Right: same as left panel, but for 
a triple configuration as in Fig.~\ref{fig:classical_laplace2}.
In general, LK oscillations are entirely suppressed for $m_p\gtrsim 2m_{p,\mathrm{crit}}$.
For smaller planet mass ($m_p< \tfrac{1}{2}m_{p,\mathrm{crit}}$), the eccentricity
oscillation amplitude is only slightly modified.
 \label{fig:maximum_ecc}}
 \vspace{-0.1in}
\end{figure*}

{In the above, our calculations have ignored the mass of the circumbinary planet $m_p$ based on the assumption
that the planetary mass is always much smaller than $M_\In$ and $M_\Out$. However, over secular
time scales, a finite planet mass can affect the dynamics of the inner binary to the point of suppressing
the eccentricity oscillations caused by the tertiary \cite{hol97}. The planet-induced precession frequency
of the binary is of order ${\Omega}_{{\text{in-}}p}\simeq n_\In\left({m_p}/{M_\In}\right)\,\left({a_\In}/{a_p}\right)^3~$.
The condition ${\Omega}_{\text{in-}p}\simeq{\Omega}_{\inout}$ allows us to define the critical planet mass
\begin{equation}\label{eq:mcrit}
m_{p,\mathrm{crit}}\simeq 0.13 M_\mathrm{J}
\left(\frac{M_{\Out}}{1M_\odot}\right)
\left(\frac{a_p}{1.5\mathrm{AU}}\right)^3
\left(\frac{a_{\Out}}{30\mathrm{AU}}\right)^{-3}
\end{equation}
(where $M_\mathrm{J}$ is the mass of Jupiter) above which the precession of the binary
due to the planet is faster that due to the tertiary star.
For small $m_p$, the effects of a finite planet mass on the LK cycles are qualitatively similar to those
of other short range forces \cite{fab07,liu15}, imposing an upper limit on the maximum eccentricity of the binary.
In Fig.~\ref{fig:maximum_ecc} we show the maximum eccentricity achieved by the inner binary 
as a function of planet mass obtained from integrations of the 4-body secular system (see Supporting materials). 
For the example depicted in Fig.~\ref{fig:smooth_transition}
 ($a_p=1.5$~AU and $m_{p,\mathrm{crit}}\simeq0.6M_\mathrm{J}$), we find that
 $m_p\gtrsim1M_\mathrm{J}\simeq1.7m_{p,\mathrm{crit}}$ is enough to
 substantially suppress the oscillations in $e_\In$. For $m_p\lesssim0.3M_\mathrm{J}\simeq0.5m_{p,\mathrm{crit}}$
 (about the mass of Saturn), the minimum pericenter separation of the binary $a_{\In,0}(1-e_{\In,\mathrm{max}})\approx0.09a_{\In,0}$
 is only $17\%$ larger than $0.077a_{\In,0}$, the value corresponding to $m_p=0$. Such a planet ($m_p\lesssim0.3M_\mathrm{J}$)
will only delay the orbital shrinkage of the inner binary, but not prevent it (see Supplementary materials for an example).}

{Throughout this paper, we have included only the quadrupole potential
from the tertiary companion acting on the inner binary and the
planet. This is a good approximation when the companion has zero
orbital eccentricity. For general companion eccentricities, octupole
and higher-order potentials may introduce more complex dynamical
behaviors for the inner binary and for the planet (see, e.g., \cite{for00,nao11,kat11,liu15}).
For
example, in $N$-body calculations (which include high order terms automatically)
the planet may attain a non-zero eccentricity as the inner binary decays
even in the moderate inclination case (see the Supplementary
material for one such example).
A systematic study of these complex ``high-order" effects is beyond the
scope of this paper and will be the subject of future work.}

\section*{Discussion}
We have explored the orbital evolution of planets around 
binaries undergoing orbital decay via the ``LK+Tide" mechanism driven by distant tertiary companions. 
We have shown that planets may survive the orbital decay of the binary
for tertiary companions at moderate initial inclinations ($i_{{\inout},0}\lesssim75^\circ$).
In such case, planets on circular orbits adiabatically follow an equilibrium solution as the triple system evolves,
becoming misaligned with their host binary;
 the final misalignment angle $i_{\pin}$ is a monotonically increasing 
function of the binary-planet distance
$a_p$.  At higher inclinations ($i_{{\inout},0}\gtrsim80^\circ$), 
the adiabatic evolution is broken when planets encounter
an unstable equilibrium. Then the planet orbit can
develop erratic behavior in eccentricity and inclination.
Very eccentric circumbinary orbits
may be disrupted by the inner binary via dynamical instabilities, resulting in either the ejection
of the planet or its collision onto the stars.
Interestingly, even in this high-inclination regime,
we have found that some planets may evolve into stable, misaligned and eccentric orbits.

In our scenario, the abundance of misaligned planets around compact binaries
depends on the frequency of moderate initial inclination stellar triples relative to those with high inclinations.
High inclination
stellar triples may be the progenitors of the majority of compact binaries, since 
the very high eccentricities reached by the inner binary make orbital decay faster. 
Our calculations suggest that planets within such high inclination triples have less chances
of survival during the inner binary's orbital decay.
The efficiency of tidal decay
depends on the dissipation time scale $t_V$ within the stars (see supplementary material).
We have found that dissipation
time scales of order $20-50$~years can circularize inner binaries with $i_{{\inout},0}\gtrsim78^\circ$ within
a Hubble time, but  if $i_{{\inout},0}\sim70^\circ$, then  $t_V\simeq1-5$~yr is required.
However, given the large parameter
space in orbital configurations, and the uncertainty in realistic values of $t_V$ (which may vary during stellar
evolution), we cannot discard the possibility that some binaries, perhaps still undergoing orbital decay and circularization,
may be part of moderate-inclination stellar triples, and may therefore be candidate hosts to highly misaligned planets.

{An additional caveat to the abundance of misaligned circumbinary planets that is not addressed in this work
concerns the likelihood of planets forming within inclined hierarchical triples with $a_\Out/a_\In\sim100$. Planet
formation will be limited by disk truncation from inside (at $a\sim3a_\In$) and from outside  (at $a\sim a_\Out/3$) 
\cite{art94,mir15}. Thus, for the parameters explored in this paper, planets would be confined to form between 
1 and 10 AU. In addition
to disk truncation and warping \cite{tre14}, planetesimal dynamics in this systems could be affected by the tidal 
forcing of the inner binary and the outer companion,
introducing additional complications to the formation of planetary cores \cite{raf15a,sil15b}.}

As noted before, currently no planets have been detected around eclipsing compact  ($P_\In\lesssim5$~days)
stellar binaries. {Our work suggests that if planets are able to form within (moderately) compact triples,
they are likely to survive the tidal shrinkage of the central binary, evolving into inclined orbits}.
The detection of these misaligned circumbinary planets may be challenging.
The planets that survive the orbital decay of the binary lie close to/on the Laplace surface, which
follows with the precession of the inner binary axis $\nvec_\In$ respect to the outer binary axis $\nvec_\Out$. 
The coupled precession of the inner binary and the planet orbits 
will produce short-lived ``transiting windows",
 but these windows appear periodically over very long time scales [of order
$1/\Omega_{\inout}\sim P_\In (a_\Out/a_\In)^3\sim 10^5-10^6$~years].
 An alternative detection strategy is to look for eclipse timing variations. {The perturbation on the inner
binary exerted by a planet of mass $m_p$ introduces a timing signature 
(on the time scale of the planet's orbital period) of the magnitude
 $\Delta P_\In\sim (3/8\pi) P_\In (m_p/M_\In)(a_\In/a_p)^{3/2}$ \cite{may90,bor03}.
 For  values of $m_p\sim 0.5 M_J\sim0.0005M_\In$, 
 $a_p\sim1.0$AU, $a_\In\sim0.05$~AU and $P_\In\sim5$~days, the maximum eclipse timing
 variation is of order  $\sim 0.3$~s, }
 approaching the noise level for some nearby binaries, but in general, still below the detection
 limits for most eclipse timing detections \cite{con14}. However, short cadence data with 
 current observational capabilities might provide enough timing precision to 
 accomplish such measurements.

\vspace{0.4in}
We thank Sarah Ballard, Konstantin Batygin, Matthew Holman and Bin Liu for discussions
and comments. 
We also thank the referee, Daniel Fabrycky for valuable comments and suggestions.
This work has been supported in part by NSF grant
AST-1211061, and NASA grants NNX14AG94G and NNX14AP31G.\\

Note added: During of the revision of our manuscript, we became aware of a preprint
by D. Martin, T. Mazeh and D. Fabrycky, which addresses a similar issue (i.e. the dearth of 
planets around compact binaries) as our paper.


\vspace{-0.3in}
\appendix
\section{}
\subsection*{S1 Equations of Motion}
Consider a binary (total mass $M_\In=m_0+m_1$, reduced mass $\mu_\In=m_0m_1/M_\In$
and semi-major axis $a_\In$)
that is a member of a hierarchical triple, in
which the binary and an outer companion of mass $M_\Out$ orbit each other
with a semi-major axis $a_\Out\gg a_\In$.
The shape and orientation of the inner binary orbit are specified by the eccentricity
vector $\evec_\In$ and the unit vector along the binary's angular momentum direction
$\nvec_\In$; similarly, the outer companion's orbit is specified by $\evec_\Out$ and $\nvec_\Out$.
A planet orbiting around the inner binary has semi-major axis $a_p$, eccentricity
vector $\evec_p$ and angular momentum direction $\nvec_p$. The perturbing potential
(per unit mass) acting on the planet has contributions from the inner binary, 
$\langle\langle\Phi_\In\rangle\rangle$,
and from the outer companion, $\langle\langle\Phi_\Out\rangle\rangle$
where the double brackets denote time averaging over the orbital periods of
the inner binary, of the outer companion and of the planet. To quadrupole order, these
potentials are given by [e.g., \cite{tre09}]
\begin{equation}\label{eq:potential_inner_av}
\begin{split}
\langle\langle\Phi_\In\rangle\rangle&=-\frac{1}{8} \frac{\mathcal{G}M_\In}{a_p}{(1-e_p^2)^{-3/2}}
\left(\frac{\mu_\In}{M_\In}\right)\left(\frac{a_\In}{a_p}\right)^2\\ 
&~~~\times\Big[1-6e_\In^2-3(1-e_\In^2)(\nvec_\In\cdot\nvec_p)^2+
15(\evec_\In\cdot\nvec_p)^2\Big]~~,
\end{split}
\end{equation}
and 
\begin{equation}\label{eq:potential_outer_av}
\begin{split}
\langle\langle\Phi_\Out\rangle\rangle&=-\frac{1}{8}\frac{\mathcal{G}M_\In}{a_p}(1-e_\Out^2)^{-3/2}
\left(\frac{M_\Out}{M_\In}\right)\,\left(\frac{a_p}{a_\Out}\right)^3\\
&~~~\times\Big[1-6e_p^2-3(1-e_p^2)(\nvec_\Out\cdot\nvec_p)^2+
15(\nvec_\Out\cdot\evec_p)^2\Big]~~.
\end{split}
\end{equation}
 In our actual calculations,
we will set the outer orbit's eccentricity $e_\Out$ to zero; this guarantees that higher order octupole terms
of the potential $\langle\langle\Phi_\Out\rangle\rangle$ are identically zero. Similarly, 
setting the inner binary to have a mass ratio of unity (which means that $\mu_\In/M_\In=1/4$)
makes the octupole terms of the inner potential $\langle\langle\Phi_\In\rangle\rangle$ vanish. 
Therefore, the quadrupole potentials given above capture the secular dynamics to high accuracy.
 \begin{figure}
\centering
\includegraphics[width=0.42\textwidth]{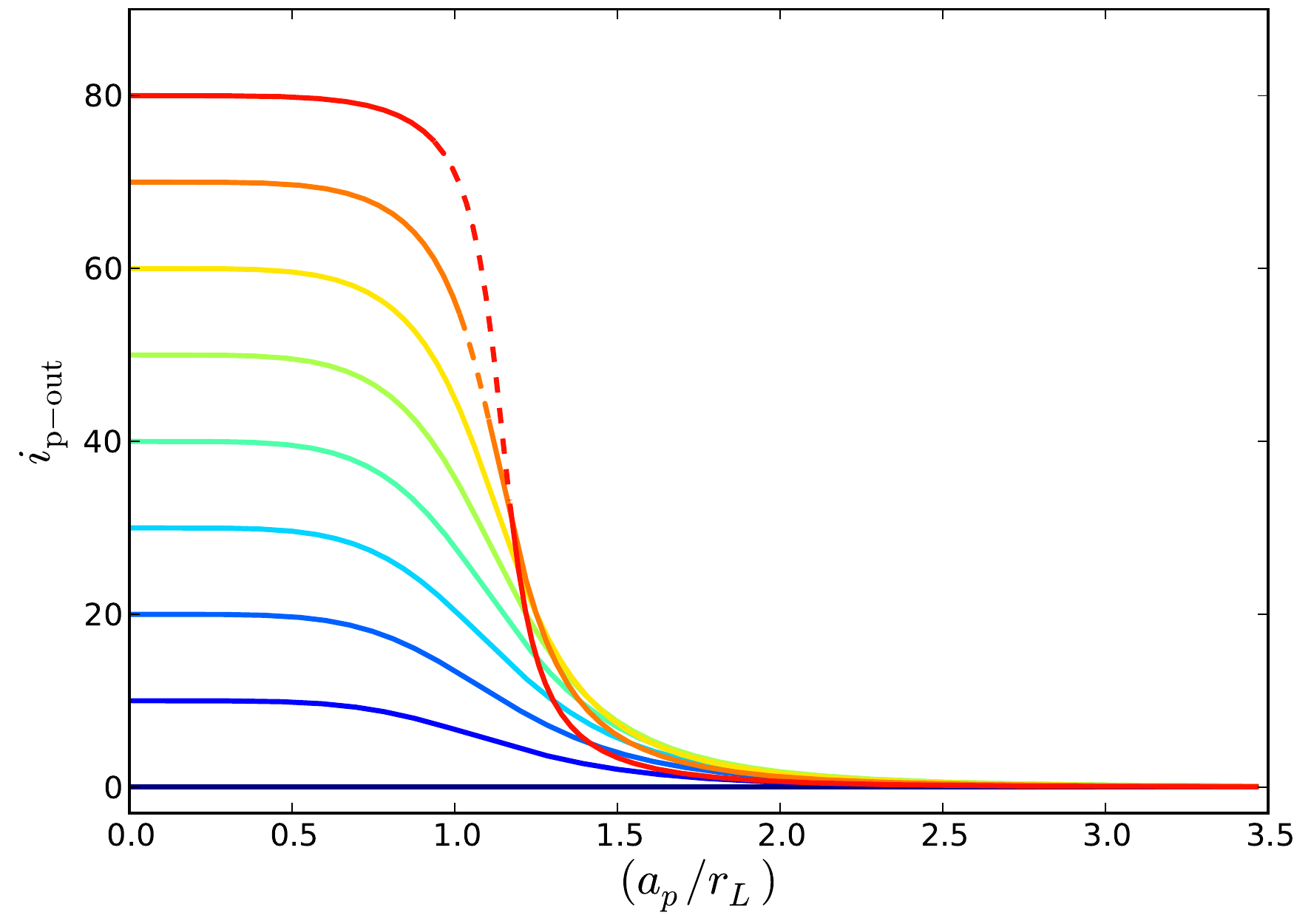}
\caption{Classical Laplace equilibrium for a test particle in a circular orbit within a hierarchical stellar triple. 
The eccentricity of the inner binary is set to zero. The inner-to-outer binary inclination $i_{\inout}$ ranges
 from $0^\circ$ (bottom blue curve) to $80^\circ$ (top red curve). The dashed portion of the curve
 indicates unstable equilibrium,
\label{fig:laplace_surface}}
\vspace{-0.1in}
\end{figure}
 
From the potentials $\langle\langle\Phi_\In\rangle\rangle$
and $\langle\langle\Phi_\Out\rangle\rangle$,  we can derive
equations of motion for the dimensionless angular momentum vector of the planet, $\jvec_p=\sqrt{1-e_p^2}\nvec_p$~,
and the eccentricity vector $\evec_p$ [see \cite{tre09,liu15}]:
 \begin{subequations}\label{eq:eqofmotion}
 \begin{align}
 \begin{split}\label{eq:jvecmotion}
\frac{d\mathbf{j}_p}{dt}=n_p
\Bigg\{
\frac{3}{2}\frac{\epsilon_\In}{(1-e_p^2)^{5/2}}
\Big[(1-e_\In^2)(\nvec_\In\cdot\mathbf{j}_p)(\mathbf{j}_p\times\nvec_\In)\\
-5(\evec_\In\cdot\mathbf{j}_p)(\mathbf{j}_p\times\evec_\In)
\Big]\\
+\frac{3}{4}\epsilon_\Out
\Big[(\mathbf{j}_p\cdot\nvec_\Out)(\mathbf{j}_p\times\nvec_\Out)
-5(\mathbf{e}_p\cdot\nvec_\Out)(\mathbf{e}_p\times\nvec_\Out)
\Big]
\Bigg\}~~,
\end{split}
\end{align}
\end{subequations}
and
\addtocounter{equation}{-1}
 \begin{subequations}
  \begin{align}
  \addtocounter{equation}{+1}
 \begin{split}\label{eq:evecmotion}
\frac{d\mathbf{e}_p}{dt}=n_p
\Bigg\{
\frac{3}{2}\frac{\epsilon_\In}{(1-e_p^2)^{5/2}}
\Big[(1-e_\In^2)(\nvec_\In\cdot\mathbf{j}_p)(\mathbf{e}_p\times\nvec_\In)~~~~~\\
-5(\evec_\In\cdot\mathbf{j}_p)(\mathbf{e}_p\times\evec_\In)
\Big]\\
-\frac{3}{4}\frac{\epsilon_\In}{(1-e_p^2)^{7/2}}\Big[
(1-6e_\In^2)(1-e_p^2)
+25 (\evec_\In\cdot\mathbf{j}_p)^2\\
-5(1-e_\In^2)(\nvec_\In\cdot\mathbf{j}_p)^2
\Big]\mathbf{j}_p\times\mathbf{e}_p\\
+\frac{3}{4}\epsilon_\Out\Big[(\mathbf{j}_p\cdot\nvec_\Out)(\mathbf{e}_p\times\nvec_\Out)
+2\mathbf{j}_p\times\mathbf{e}_p\\
-5(\mathbf{e}_p\cdot\nvec_\Out)(\mathbf{j}_p\times\nvec_\Out)\Big]
\Bigg\}~~,
\end{split}
\end{align}
 \end{subequations}
 %
 \begin{figure*}
\centering
\includegraphics[width=0.45\textwidth]{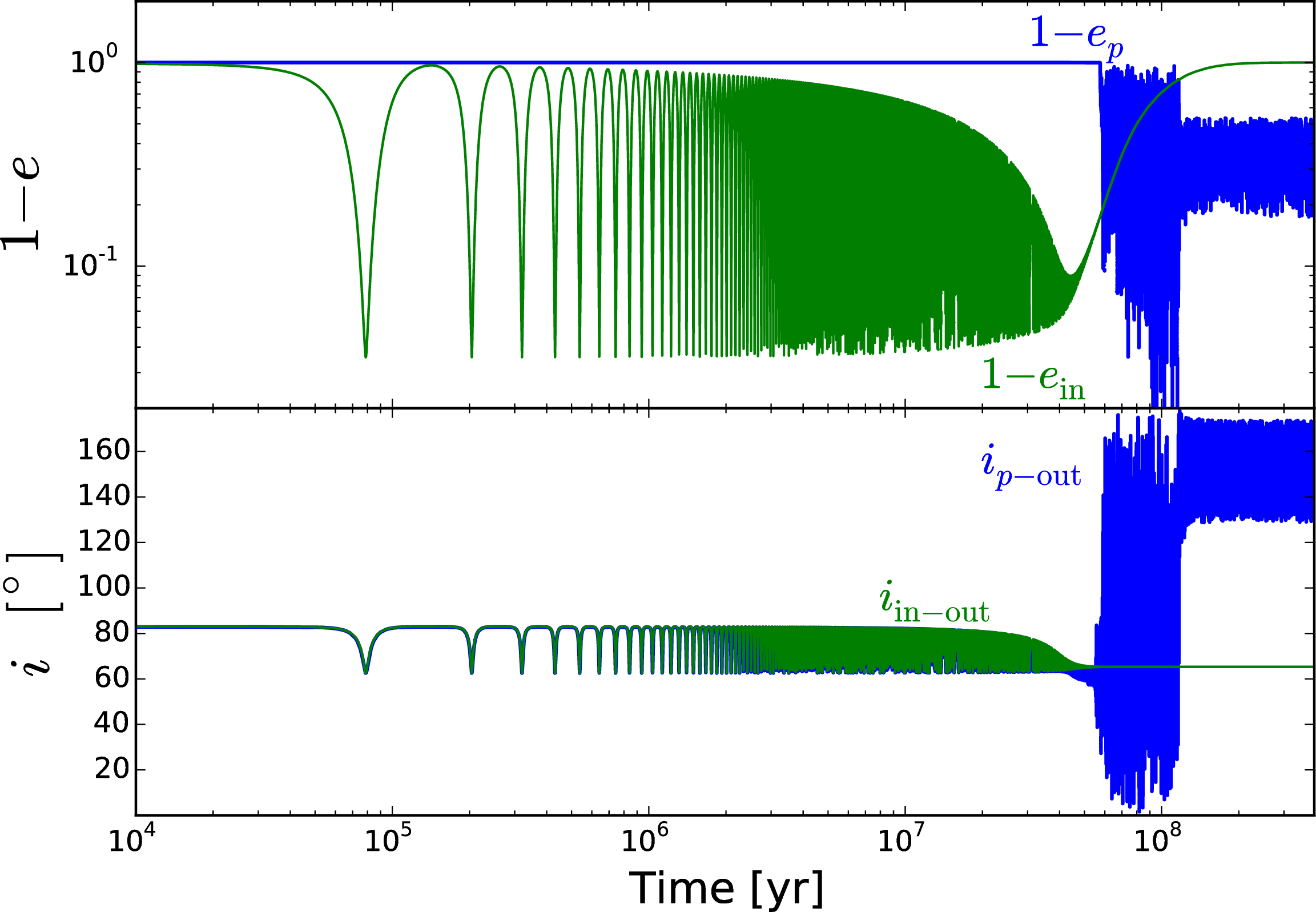}
\includegraphics[width=0.45\textwidth]{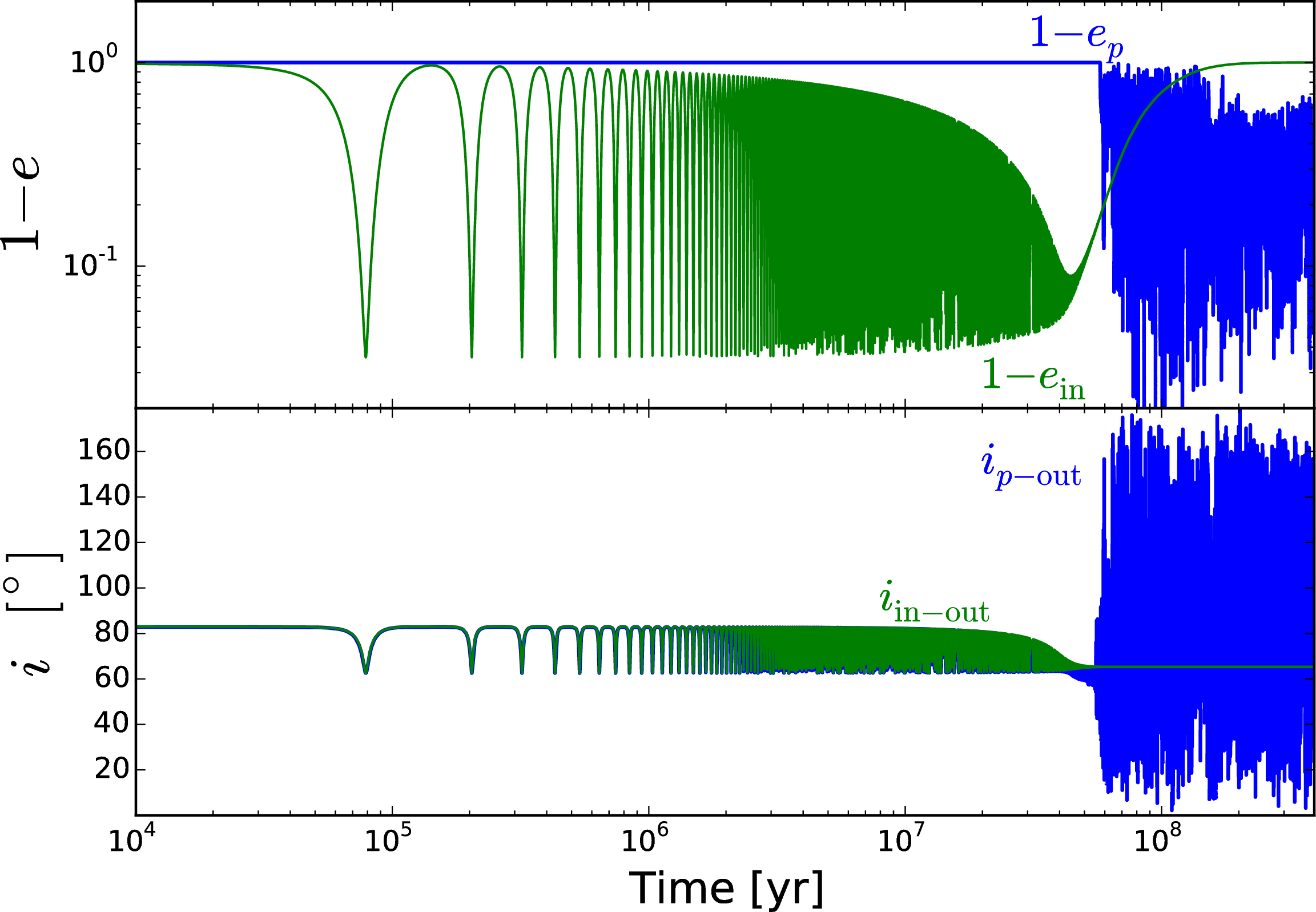}
\caption{Two examples of planet orbital evolution with nearly identical initial conditions showing
very different outcomes. As in Fig.~5 in the main text,  the triple system parameters are 
 $a_{\In,0}=0.3$~AU,
$a_\Out=30$~AU and $i_{\inout,0}=83^\circ$. The left panels show the eccentricity and inclination 
evolution of the 
planet (blue) and inner binary (green) when $a_p=1.6000$~AU. The right panels show the same
quantities (the evolution of the inner binary is identical in both cases) for a planet with $a_p=1.6002$~AU. 
The dramatic difference between the two outcomes of the planet's eccentricity and inclination
 indicates the chaotic nature of planetary orbits undergoing erratic evolution
for high-inclination stellar triples.
\label{fig:chaotic}}
\end{figure*}
where we have introduced the dimensionless quantities
\begin{subequations}
\begin{align}
\epsilon_\In&\equiv \frac{1}{2}\left(\frac{\mu_\In}{M_\In}\right)\left(\frac{a_\In}{a_p}\right)^2~~,\\
\epsilon_\Out&\equiv  \frac{1}{(1-e_\Out^2)^{3/2}}\left(\frac{M_\Out}{M_\In}\right)\left(\frac{a_p}{a_\Out}\right)^3~~.
\end{align}
\end{subequations}
Note that, in the notation of the main body of this article $\Omega_{\pin}=n_p \epsilon_\In$ and 
$\Omega_{\pout}=n_p \epsilon_\Out$.

By setting $d\jvec_p/dt=d\evec_p/dt=0$ we obtain the equilibrium solution for the planet's orbit under the influence of both
inner and outer torques. This is known as  ``classical Laplace equilibrium" in the special case of
$e_\In=0$ \cite{tre09,tam13}. From Eq.~(\ref{eq:jvecmotion}), and setting $e_\In=0$,
the equation for coplanar Laplace equilibrium reads:
 \begin{equation}
0=2\epsilon_\In
(\nvec_\In\cdot\mathbf{j})(\mathbf{j}\times\nvec_\In)
+\epsilon_\Out(\mathbf{j}\cdot\nvec_\Out)(\mathbf{j}\times\nvec_\Out)~~.
 \end{equation}
Using orbital elements, we can write this expression as \cite{tre09,tam13}
 \begin{equation}\label{eq:laplace_surface2}
0=2\epsilon_\In \sin2(i_{\pout}-i_{\inout})+\epsilon_\Out\cos2i_{\pout}~~.
 \end{equation}
 We can solve this transcendental equation for $i_{\pout}$ as a function of the ratio
 $\epsilon_\Out/\epsilon_\In$, or equivalently, as a function of the Laplace radius $r_L$ 
 (see Eq.~5 in the main body of this article), since $(r_L/a_p)^5=\epsilon_\In/\epsilon_\Out$. 
 Fig.~\ref{fig:laplace_surface} shows the Laplace surface solution as a function of $r_L/a_p$
for different values of $i_{\inout}$ ranging from $0^\circ$ to $90^\circ$. This solution of
the classical Laplace surface is used in Figures~4 and 5 of the main body of the article.

In our numerical calculations, we directly integrate Eqs.~\ref{eq:eqofmotion}
together with the evolution equations of the hierarchical triple as the inner binary
undergoes LK oscillations with tidal dissipation. In principle, this 
system consists of 24 coupled differential equations (involving the vectors
$\jvec_\In$, $\evec_\In$, $\jvec_p$, $\evec_p$, $\jvec_\Out$, $\evec_\Out$ for the orbits,
and the spin vectors $\boldsymbol{\Omega}_0$ and $\boldsymbol{\Omega}_1$ for each of the
two central stars). We have simplified this system as follows: neglect
the evolution of the outer orbit (valid approximation when the outer angular momentum dominates);
include only short range forces
acting on the secondary central star (i.e. the primary star is a non-spinning ``rigid sphere");  
assume psedo-synchronization [e.g., (16)] and orbital alignment for the spin 
vector $\boldsymbol{\Omega}_1$. This simplification reduces the number of equations
to 12, while still capturing all the essential physics of the problem.

 \begin{figure*}
 \begin{center}
\includegraphics[width=0.47\textwidth]{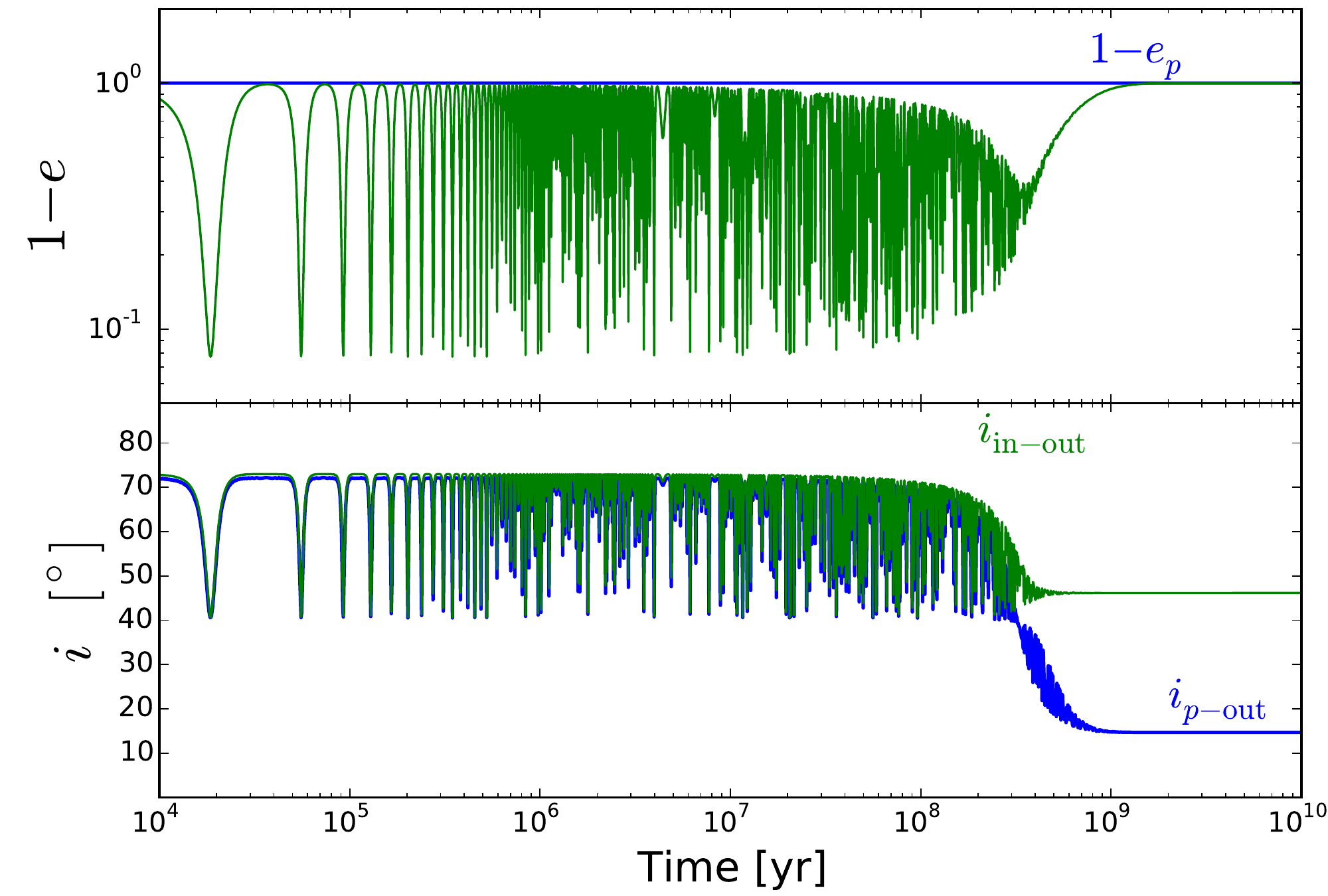}
\includegraphics[width=0.47\textwidth]{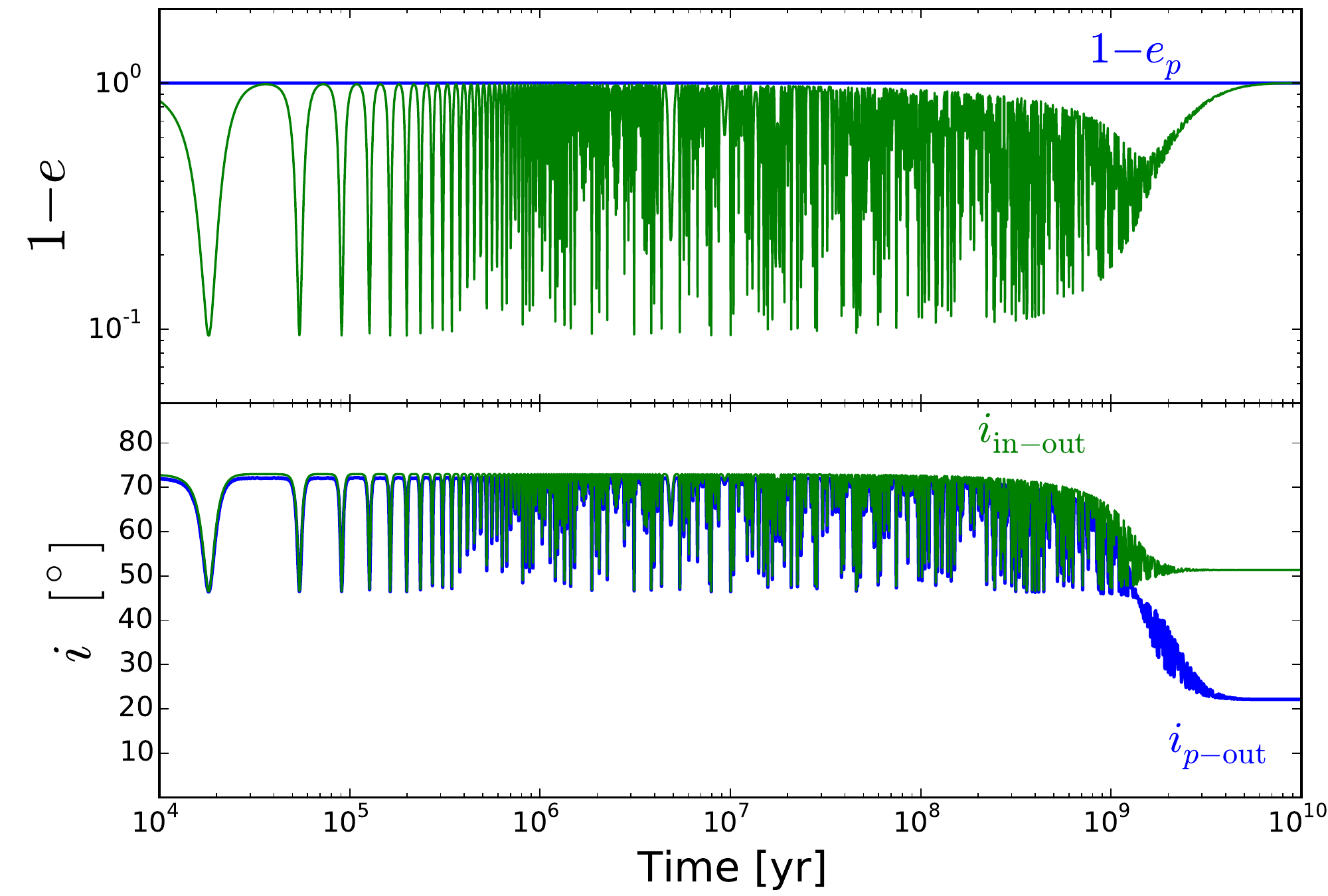}
\caption{Effects of planet mass on the orbital evolution of the inner binary and the planet.
Left: Orbital evolution of a binary ($a_{\In,0}=0.3$~AU) and a circumbinary planet  ($a_p=1.5$~AU)
in a triple with $i_{\inout,0}=73^\circ$ and $a_\Out=18$~AU
where $m_0+m_1=M_\In=M_\Out=1M_\odot$ and $m_p=1$ Earth mass. The circularization of the inner binary
takes place at the same rate as the case with $m_p=0$. Right: Same as the left panel, but with $m_p=1$ Saturn mass.
Circularization is not prevented, but it takes $\sim4$ times longer than the $m_p=0$ case. 
\label{fig:massive_planet}}
\end{center}
\end{figure*}

For the evolution of the  vectors $\evec_\In$ and $\nvec_\In=\jvec_\In/|\jvec_\In|$,  we use the formalism
of  \cite{egg01} and \cite{fab07}:
 \begin{subequations}\label{eq:eqofmotion_inner}
 \begin{align}
\frac{d\mathbf{j}_\In}{dt}=
\frac{d\mathbf{j}_\In}{dt}\Big|_{\inout}+\frac{d\mathbf{j}_\In}{dt}\Big|_{\mathrm{SRF}}\\
\frac{d\mathbf{e}_\In}{dt}=
\frac{d\mathbf{e}_\In}{dt}\Big|_{\inout}+\frac{d\mathbf{e}_\In}{dt}\Big|_{\mathrm{SRF}}
\end{align}
\end{subequations}
where the first terms on the right  hand side correspond to  the conservative tidal effect of
the tertiary  and the second terms to the conservative and non-conservative short range forces.
In ${d\mathbf{j}_\In}/{dt}\big|_{\mathrm{SRF}}$ and ${d\mathbf{e}_\In}/{dt}\big|_{\mathrm{SRF}}$,
an important parameter is the
viscous time $t_V$ associated with the dissipation within the stars.  This parameter
is the main source of uncertainty in our calculations. We vary $t_V$ between
 $\sim1$ year to 55 years [the value used by \cite{fab07}]. Note that
we include dissipation only within one of the stars. 
The effective tidal quality factor $Q$ scales proportionally with $t_V$ as \cite{egg01}:
\begin{equation}
Q=\frac{4}{3}\frac{k}{(1+2k)^2}\frac{\mathcal{G}m_1}{R_{*,1}^3} \frac{t_V}{n_\In}~~,
\end{equation}
where we use $m_1=m_0=0.5M_\odot$, $R_{*,1}=R_{*,0}=0.5R_\odot$ and $k$ (the classical apsidal
motion constant) is set to $0.014$.  We see that one to two orders of magnitude of variation
 in $t_V$ is not unreasonable,
given the large degree of uncertainty in the tidal $Q$ values for different types of stars.
The assumption
of pseudo-synchronization and spin-orbit alignment also introduces uncertainties, although
the dominant uncertainty still lies in the value of $t_V$. These uncertainties ultimately affect not the final, 
circularized orbit of the inner binary, but how fast this
final state can be reached (see the discussion in the main body of the text).

\subsection*{S2 Erratic evolution and chaotic behavior}
In Fig.~\ref{fig:chaotic} we show two integrations for planets within the stellr triple configuration of 
Fig.~6 in the main text.
The left panels show the eccentricity and inclination evolution of a planet with $a_p=1.6$~AU, and the right
panels show the same for a planet with $a_p=1.6002$ (about a $0.1\%$ difference in semi-major axis).
The evolution of both planets is identical until circular orbits become unstable, at which point rapid eccentricity
growth is triggered, followed by highly erratic behavior in the evolution of $e_p$ and $i_{\pout}$. The first example ``exits" the
erratic region and finds a stationary state in which $e_p$ and $i_{\pout}$ oscillate around a well-defined
value with constant amplitude (in this case, the planet lands on a retrograde orbit respect to $\nvec_\Out$).
However, the second example never finds ``a way out" of the erratic region (which
must happen before the inner binary circularizes, ``locking" the properties of the system).
Since both these examples go through a phase of extreme eccentricities, they are both likely to be ejected by
interactions with the inner binary, and thus their chances of surviving for a long period of time are equally small, regardless
of the duration of the erratic phase.

\subsection*{S3 Effect of finite planet mass}

In the calculations above, the mass of the planet $m_p$ has been neglected.  However,
the effect of finite planet mass on the inner binary can be taken into account in a self-consistent
fashion by including the tidal potential on the inner binary
due to the planet, which introduces the additional terms  
${d\mathbf{j}_\In}/{dt}\big|_{\text{in-}p}$ and ${d\mathbf{e}_\In}/{dt}\big|_{\text{in-}p}$ to
Eq.~\ref{eq:eqofmotion_inner}. These extra 
 terms have the same functional form
as the terms arising from the tidal potential of the tertiary
${d\mathbf{j}_\In}/{dt}\big|_{\text{in-out}}$ and ${d\mathbf{e}_\In}/{dt}\big|_{\text{in-out}}$.
Fig.~\ref{fig:massive_planet} depicts a similar example to that of Fig.~3, this time
including the effects of $m_p\neq0$. For a planet of 1 Earth mass, the results are indistinguishable
from the test-particle case. However, for a planet of 1 Saturn mass, the effects of the modified
maximum binary eccentricity (see Fig.~7) can be readily seen in the efficiency at which the binary shrinks.

 \begin{figure}
 \begin{center}
\includegraphics[width=0.45\textwidth]{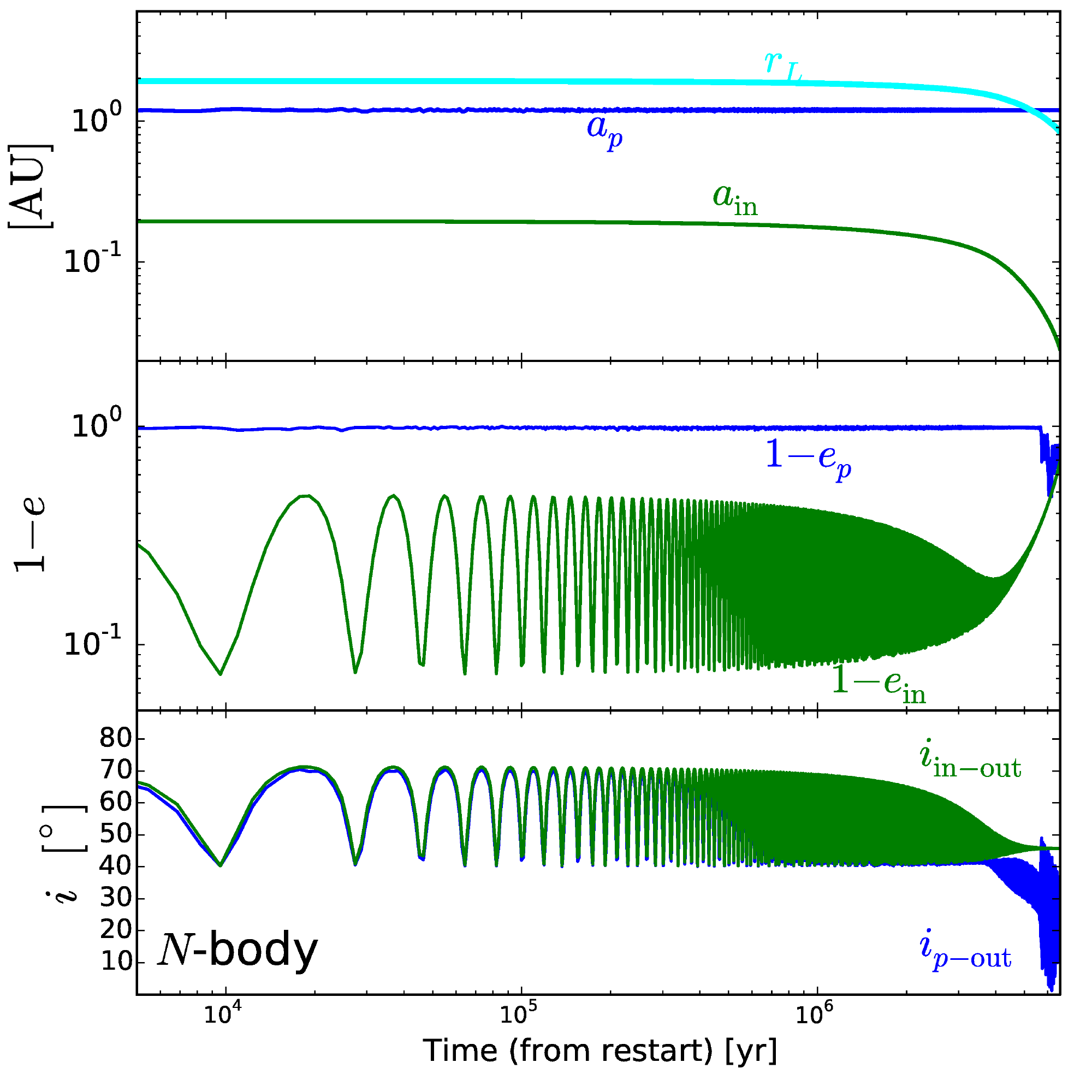}
\vspace{-0.05in}
\caption{Example of a dissipative $N$-body integration of a 4-body system.
Masses of the binary and companion are $m_0=m_1=0.5M_\odot$ and $M_\Out=1M_\odot$.
The companion semimajor axis is $a_\Out=18$~AU and its eccentricity is $0.2$. The planet
semimajor axis is $a_p=1.2$~AU and its mass is zero. 
The $N$-body integration is started once the binary has undergone appreciable orbital decay,
when $a_\In=0.19$~AU, $e_\In=0.516$ and $i_{\inout}=73^\circ$. The line types are
the same as in Fig.~3.
\label{fig:nbody}}
\vspace{-0.22in}
\end{center}
\end{figure}

\subsection*{S4 Example of an $N$-body integration}
Although long-term direct $N$-body integrations are computationally costly, one can
combine them with the output of the secular solutions to study the detailed behavior
of the planetary orbit as the binary decays. 
Fig.~\ref{fig:nbody} shows the $N$-body result of a dissipative binary 
performed using the {\footnotesize MERCURY} code \cite{cha99}. We have added short range
forces following \cite{bea12}, including stellar tides, GR and tidal dissipation. 
This integration is started once the binary has undergone appreciable orbital decay, 
but before the planet has crossed the Laplace radius $r_L$;
the initial condition is obtained using the results of our secular calculation. 
In our example, the orbital evolution has been accelerated by increasing the tidal dissipation rate.

As seen from Fig.~\ref{fig:nbody}, before crossing the Laplace radius, the planet's inclination
adiabatically follows that of the inner binary while maintaining zero eccentricity, in full
agreement with our secular results. After $a_p$ crosses $r_L$, the planet develops a modest
eccentricity and its inclination decouples from that of the binary.  Note that since the tertiary
companion has a finite eccentricity (0.2) in this example, the octupole potential on the planet
likely plays an important role. This may explain the development of finite planet eccentricity
for a system with such a modest $i_{\inout}$.


\end{document}